\documentclass[%
 aip,
jcp,
 amsmath,amssymb,
preprint,%
]{revtex4-1}

\usepackage[english]{babel}

\usepackage{graphicx}
\usepackage{amssymb,amsmath}

\usepackage{color}

\begin{document}

\title{Extended law of corresponding states for protein solutions}

\author{Florian Platten}
\email[]{Florian.Platten@hhu.de}
\affiliation{Condensed Matter Physics Laboratory, Heinrich Heine University, 40225 D\"usseldorf, Germany}

\author{N\'estor E. Valadez-P\'erez}
\affiliation{Division of Sciences and Engineering, University of Guanajuato, 37150 Le\'on, Mexico}

\author{Ram\'on Casta\~neda-Priego}
\affiliation{Division of Sciences and Engineering, University of Guanajuato, 37150 Le\'on, Mexico}

\author{Stefan U. Egelhaaf}
\affiliation{Condensed Matter Physics Laboratory, Heinrich Heine University, 40225 D\"usseldorf, Germany}

\begin{abstract} 
The so-called extended law of corresponding states, as proposed by Noro and Frenkel [J. Chem. Phys. \textbf{113}, 2941 (2000)], involves a mapping of the phase behaviors of systems with short-range attractive interactions.
While it has already extensively been applied to various model potentials, here we test its applicability to protein solutions with their complex interactions. 
We successfully map their experimentally determined metastable gas--liquid binodals, as available in the literature, to the binodals of short-range square-well fluids, as determined by previous as well as new Monte Carlo simulations.
This is achieved by representing the binodals as a function of the temperature scaled with the critical temperature (or as a function of the reduced second virial coefficient) and the concentration scaled by the cube of an effective particle diameter, where the scalings take into account the attractive and repulsive contributions to the interaction potential, respectively.
The scaled binodals of the protein solutions coincide with simulation data of the adhesive hard-sphere fluid.
Furthermore, once the repulsive contributions are taken into account by the effective particle diameter, the temperature dependence of the reduced second virial coefficients follows a master curve that corresponds to a linear temperature dependence of the depth of the square-well potential.
We moreover demonstrate that, based on this approach and cloud-point measurements only, second virial coefficients can be estimated, which we show to agree with values determined by light scattering or by DLVO-based calculations.
\end{abstract}

\maketitle



\section{Introduction}
\label{sec:1}
Proteins show a rich phase behavior; they can crystallize,\cite{Durbin1996,Doye2006} undergo liquid--liquid phase separation (LLPS),\cite{Broide1996,Muschol1997} or form aggregates\cite{DeYoung1993,Chi2003} or fibrils.\cite{Dobson2003,Evers2009b,Seeliger2012} 
It depends on the protein--protein interactions which can be tuned through the solution conditions, namely thermodynamic parameters, such as temperature, \textit{p}H, salt and protein concentration.\cite{Vekilov2012,Mezzenga2013} 
Controlling the phase behavior of protein solutions is of fundamental importance in medicine, materials science, physics and several other disciplines, as illustrated by the following examples: 
Condensation of protein phases, as observed in LLPS and fibril formation, plays a major role in the pathogenesis of several human diseases,\cite{Vekilov2012,Mezzenga2013} such as cataract formation and Alzheimer's disease.\cite{Siezen1985,Chiti2006} 
Unwanted protein phase transitions can severely impede drug delivery and purification in pharmaceutical treatments and food engineering.\cite{Shukla2011,Kamerzell2011,Mezzenga2013} 
Moreover, conditions under which high quality protein crystals grow, which are an essential prerequisite for diffraction studies, are related to the protein phase behavior.\cite{Wolde1997} 

Protein solutions represent a highly complex system comprising protein molecules, water and usually also salts and further additives. 
Hence, there are a number of different contributions to intermolecular interactions among proteins, such as electrostatic interactions, van der Waals forces, hydrophobic interactions and hydration, ion-dispersion forces, and ionic or other bridges, etc., which are often highly directional.
Due to this complexity, the protein--protein interaction potential $U(\vec r)$, where $\vec r$ is the center-to-center vector of protein molecules, is usually not known in detail. 
Nevertheless, various models from liquid-state theory and colloid science, in which protein solutions are represented as effective one-component systems, have successfully been applied to study protein phase behavior. 
In many cases, protein--protein interactions are modelled by a hard-core repulsion, a short-range attraction and a long-range repulsion. 
For example, the Derjaguin-Landau-Verwey-Overbeek (DLVO) potential consists of a hard core, a short-range attractive van der Waals contribution and screened Coulomb interactions, thus directly taking into account the charges of proteins and charge screening due to salts. 
Several authors have studied protein phase behavior using DLVO-based models.\cite{Muschol1995,Velev1998,Pellicane2004a,Pellicane2012} 
Applying modified DLVO theory, salt concentration and \textit{p}H effects \cite{Poon2000,Sedgwick2005} as well as the effect of glycerol \cite{Sedgwick2007,Goegelein2012} on protein crystallization could be explained. 
Although DLVO is not able to explain, e.g., salt-specific effects,\cite{Piazza1999,Ninham1999} it has been argued \cite{Pellicane2004} that the functional form of the (attractive part of) DLVO potential is flexible enough to also account for non-DLVO forces, such as hydration, if the interaction parameters (e.g., the Hamaker constant) are suitably adjusted. 
Especially in Monte Carlo simulations, the hard-core attractive Yukawa (HCAY) \cite{Rosenbaum1996,Rosenbaum1999,Hagen1994,Dumetz2008,Valadez-Perez2012} and the Asakura-Oosawa (AO) potential \cite{Kulkarni1999,Kulkarni2001} have also been applied to model protein--protein interactions in the presence of additives. 
The simplest model accounting for attractive forces is the square-well (SW) fluid. 
short-range SW potentials have frequently been used to model protein--protein interactions,\cite{Asherie1996,Lomakin1996,Grigsby2001,Pagan2005,Liu2007,Duda2009,Valadez-Perez2012} including the analysis of small-angle scattering data of protein solutions.\cite{Zhang2007}
Moreover, Baxter's model, i.e., the adhesive hard-sphere (AHS) potential, which represents a specific limit of the SW potential, has also been applied to study protein--protein interactions.\cite{Piazza1998,Rosenbaum1996,Rosenbaum1996a} 
Recently, simulation work on anisotropic interactions of patchy particles accounting for the high directionality of protein interactions was successfully compared with binodals of protein solutions.\cite{Sear1999,Kern2003,Liu2007,Foffi2007,Goegelein2008,Fusco2013,Roosen-Runge2014} 
Although the effective interaction models \cite{Likos2001,Wolf2014} involve a (crude) simplification of the interactions in protein solutions, reasonable modelling of protein phase behavior is achieved if the short-range nature of the interactions is considered.

An integral parameter, which is independent of the specific shape of the interaction potential, is the second virial coefficient $B_2$. 
It provides a quantitative measure of interactions: positive (negative) values of $B_2$ indicate net repulsive (attractive) interactions. 
For interaction potentials with spherical symmetry, $U(r)$, the second virial coefficient is given by
\begin{eqnarray}
B_2 = 2 \pi \int_0^\infty \left( 1 - \exp{\left[-\frac{U(r)}{k_\text{B} T} \right]} \right) r^2 \text{d}r ,
\end{eqnarray}
where $k_\text{B}$ is Boltzmann's constant and $T$ the absolute temperature. 
The normalized second virial coefficient $b_2$ is defined as    
\begin{eqnarray}
b_2 = \frac{B_2}{B_2^{\text{HS}}},
\end{eqnarray}
where $B_2^{\text{HS}} = \frac{2}{3}\pi \sigma^3$ is the second virial coefficient of a hard sphere system with particle diameter $\sigma$. 
Second virial coefficients can be determined experimentally, e.g., by scattering techniques and osmometry.\cite{Goegelein2012,Gibaud2011,Rosenbaum1999,Moon2000} 

Some empirical ``laws'' linking the phase behavior of colloids and proteins to the second virial coefficient have been established.
Based on experiments, George and Wilson \cite{George1994} have shown that optimum solution conditions for the crystallization of a number of proteins are characterized by a range of $b_2$ values, the so-called crystallization slot:\cite{George1994,Vliegenthart2000}
\begin{eqnarray}
-10 \lesssim b_2 \lesssim -1.
\end{eqnarray}
For lower $b_2$ values, attractions between protein molecules are so strong, that amorphous protein aggregation occurs, while, for larger $b_2$ values, repulsions dominate and crystallization is too slow or impossible.
 
For fluids with short-range attractions, quantitative scaling laws for the gas--liquid binodal near the critical point have been proposed based on simulation results. 
Vliegenthart and Lekkerkerker (VL) \cite{Vliegenthart2000} have found that, for many model potentials, the second virial coefficient takes an approximately constant value at the critical temperature $T_\text{c}$:
\begin{eqnarray} \label{eq:vl}
b_2 (T_\text{c}) = b_2^\text{c} \approx -1.5 .
\end{eqnarray}
Moreover, Eq.~(\ref{eq:vl}) can be used as a predictor for the critical temperature: if, for a given temperature $T$, one finds $b_2 (T) \approx -1.5$, then Eq.~(\ref{eq:vl}) implies $T \approx T_\text{c}$. 
Recently, Wolf et al. \cite{Wolf2014} confirmed the VL finding. 
They argue that the osmotic pressure of a protein solution close to its critical point is low and hence it can be expanded into a virial series truncated after third order in density, yielding an approximate relationship between the critical volume fraction $\phi_\text{c}$ and second virial coefficient $b_2^\text{c}$ at the critical temperature:
\begin{eqnarray}\label{eq:1}
b_2^\text{c} \approx -\frac{1}{4 \phi_\text{c}} ,
\end{eqnarray}  
which, based on Eq.~(\ref{eq:vl}), suggests $\phi_\text{c} \approx 0.17$ in agreement with other findings (e.g., Sec.~\ref{sec:2}, Fig.~\ref{fgr:1}c).

Applying the Weeks-Chandler-Andersen (WCA) method,\cite{Weeks1971,Andersen1971} according to which an interaction potential $U(r) = U_{\text{rep}}(r) + U_{\text{attr}}(r)$ is separated into a contribution containing all repulsive interactions, $U_{\text{rep}}(r)$, and one containing all the attractive parts of the potential, $U_{\text{attr}}(r)$, Noro and Frenkel (NF) \cite{Noro2000} have suggested that many model potentials with sufficiently short-range attractions can be characterized by only three quantities: (i) an effective hard-core diameter $\sigma_\text{eff}$ taking into account all repulsive interactions between the particles,\cite{Barker1976} which is given by the relationship
\begin{eqnarray} \label{eq:eff}
\sigma_{\text{eff}} = \int_0^\infty \left( 1 - \exp{\left[ -\frac{U_{\text{rep}}(r)}{k_\text{B} T} \right]} \right) \text{d}r ,
\end{eqnarray}
(ii) an energy scale $\epsilon$ being a measure for the strength of the attraction, and (iii) the reduced second virial coefficient $b_2^\star$, which reads
\begin{eqnarray} \label{eq:4}
b_2^\star = \frac{B_2}{B_2^{\text{HS}\star}} = \frac{B_2}{\frac{2}{3}\pi \sigma_{\text{eff}}^3}  .
\end{eqnarray}
It is related to the range (and strength) of the attractions only, because the contributions of the repulsive interactions to $B_2$ are balanced by considering $\sigma_\text{eff}$ instead of $\sigma$.
Based on their mapping onto an equivalent square-well fluid, Noro and Frenkel proposed an extended law of corresponding states (ELCS): 
Many colloidal systems with the same reduced temperature $T^\star = k_\text{B} T / \epsilon$, density $\rho^\star = \rho \sigma_\text{eff}^3$ (where $\rho$ is the particle number density), and second virial coefficient $b_2^\star$ obey the same equation of state, such that the compressibility factor $z$ is a function of only three parameters: 
\begin{eqnarray}
z=\frac{\Pi}{\rho k_\text{B}T}=z(T^\star,\rho^\star,b_2^\star),
\end{eqnarray}  
where $\Pi$ is the osmotic pressure of the system.

While for many model potentials, including HCAY, AO and patchy particles, the applicability, possible extensions and restrictions of the ELCS have been studied by simulations,\cite{Foffi2007,Orea2008,Valadez-Perez2012,Gazzillo2013} the ELCS has hardly been assessed for systems as complex as proteins.\cite{Gibaud2011} 
Though it has been found that scaling gas--liquid binodals of protein solutions to the critical points results in a master curve,\cite{Katsonis2006,Duda2009} most experimental values of the second virial coefficient $b_2^\text{c}$ are systematically smaller than those predicted by VL and NF.\cite{Rosenbaum1999,Gibaud2011,Goegelein2012} 
To our knowledge, systematic attempts to clarify the differences between the quantitative predictions of the ELCS and experimental data of the gas--liquid binodal of protein solutions and the corresponding temperature dependent second virial coefficients have not been undertaken yet. 

In this work, we aim for a direct and quantitative comparison of the predictions of the ELCS with the experimentally observed phase behavior of protein solutions, namely the metastable gas--liquid binodal, and the temperature dependent second virial coefficient. 
This comparison is based on experimental data available in the literature that provide both the phase behavior and second virial coefficients and also on previous and new Monte Carlo (MC) data on short-range SW fluids.
We start (Sec.~\ref{sec:2}) by applying the ELCS to short-range SW, AHS, HCAY and AO fluids.
In the $b_2 - \phi$ plane (where $\phi$ is the particle volume fraction), the binodals of all these fluids match.
Then, in Sec.~\ref{sec:3}, the experimentally determined binodals of protein solutions are compared with those of SW fluids.
Including the repulsive part of the protein--protein interactions through an effective hard-core diameter $\sigma_\text{eff}$ yields a reasonable collapse of the experimental protein data onto a single binodal in the $T / T_\text{c} - \phi_\text{eff}$ plane (where $\phi_\text{eff}$ is the effective volume fraction based on $\sigma_\text{eff}$).
In Sec.~\ref{sec:4}, we subsequently investigate the temperature dependence of the second virial coefficient.
If the effective hard-core diameter $\sigma_\text{eff}$ is taken into account, for a variety of solution conditions close to the critical temperature $T_\text{c}$, the experimental values of the reduced second virial coefficient $b_2^\star (T / T_\text{c})$ are found to lie near a master curve, that corresponds to a linear temperature dependence of the depth of the SW potential.
In Sec.~\ref{sec:5}, then the $b_2^\star (T / T_\text{c})$ values are used to plot the experimental protein binodals in the $b_2^\star-\phi_\text{eff}$ plane.
They match with the binodals of the short-range SW and the AHS fluids, thus confirming the applicability of the ELCS to protein solutions.
Furthermore, in Sec.~\ref{sec:6}, we show that, based on the DLVO potential, the $b_2$ values of protein solutions can successfully be calculated for different salt concentrations and $p$H values.
However, using DLVO theory the effective hard-core diameter is underestimated, indicating that the protein--protein potential contains repulsive contributions beyond the electrostatic repulsion.
We finally demonstrate in Sec.~\ref{sec:7} how our approach can be exploited to determine $b_2$ values based on cloud-point measurements only, which are shown to quantitatively agree with values obtained by light scattering experiments or by calculations based on the DLVO potential.
We conclude by summarizing our findings in Sec.~\ref{sec:8}.


\section{Square-well fluids and the ELCS for simple fluids}\label{sec:2}
The square-well (SW) potential has a simple mathematical structure and provides an unambiguous definition of the interaction range. 
It is often used as a simple model for the interactions governing the phase behavior of protein and colloid systems. 
In order to provide a basis for the mapping of the protein phase behavior onto an equivalent SW potential, following Noro and Frenkel,\cite{Noro2000} we discuss features of short-range SW fluids in this section.
This discussion is based on our new MC simulation data as well as results available in the literature.

The coexistence of two fluid phases, a gas phase and a liquid phase, in a one-component system is only possible if attractive particle--particle interactions are present. 
The simplest model showing this scenario is probably given by the square-well (SW) potential $U_\text{SW}(r)$.
It consists of a hard-core repulsion of range $\sigma$ (the diameter of the particle), which leads to excluded volume effects, and a constant attractive part, which extends to a distance $\lambda \sigma$ from the center. 
Thus, the SW potential is mathematically defined as
\begin{eqnarray}
U_\text{SW}(r) = \begin{cases}\infty, \quad & r < \sigma, \\-\epsilon, \quad & \sigma \leq r \leq \lambda \sigma, \\0, \quad & r > \lambda \sigma, \end{cases} 
\end{eqnarray}
where $\epsilon > 0$ is the depth of the square well, quantifying the strength of attraction.\cite{Schoell-Paschinger2005} 
(Note that the dimensionless width of the SW, $\delta = \lambda - 1$, is also used in the literature.)

The phase behavior of square-well fluids is governed by the ratio of the range $\lambda \sigma$ and the particle size $\sigma$.\cite{Miller2004} 
It has been shown\cite{Asherie1996,Liu2005a,Pagan2005} that for $\lambda \lesssim 1.25$ gas--liquid phase separation is metastable with respect to the fluid--solid equilibrium. 
We will refer to SW fluids with $\lambda \lesssim 1.25$ as short-range SW fluids. 
Simulations indicate that also for a Lennard-Jones potential metastable gas--liquid phase separation occurs for similar values of the range.\cite{Wolde1997} 
Experiments and theory on colloid-polymer mixtures \cite{Ilett1995,Lekkerkerker1992} have shown that the topology of the phase diagram changes when the range of interaction is about $\lambda \approx 1.25$.
Moreover, the limit of short ranges $\lambda$ is relevant for exploring the relation of the SW potential to the adhesive hard-sphere (AHS) potential,\cite{Baxter1968,Watts1971,Miller2003,Miller2004,Miller2004a,Largo2008} which is defined by an infinitesimal interaction range $\lambda$ in combination with an infinite interaction strength such that $b_2$ remains finite. 

\begin{figure}
  	\centering
   \includegraphics[width=18cm]{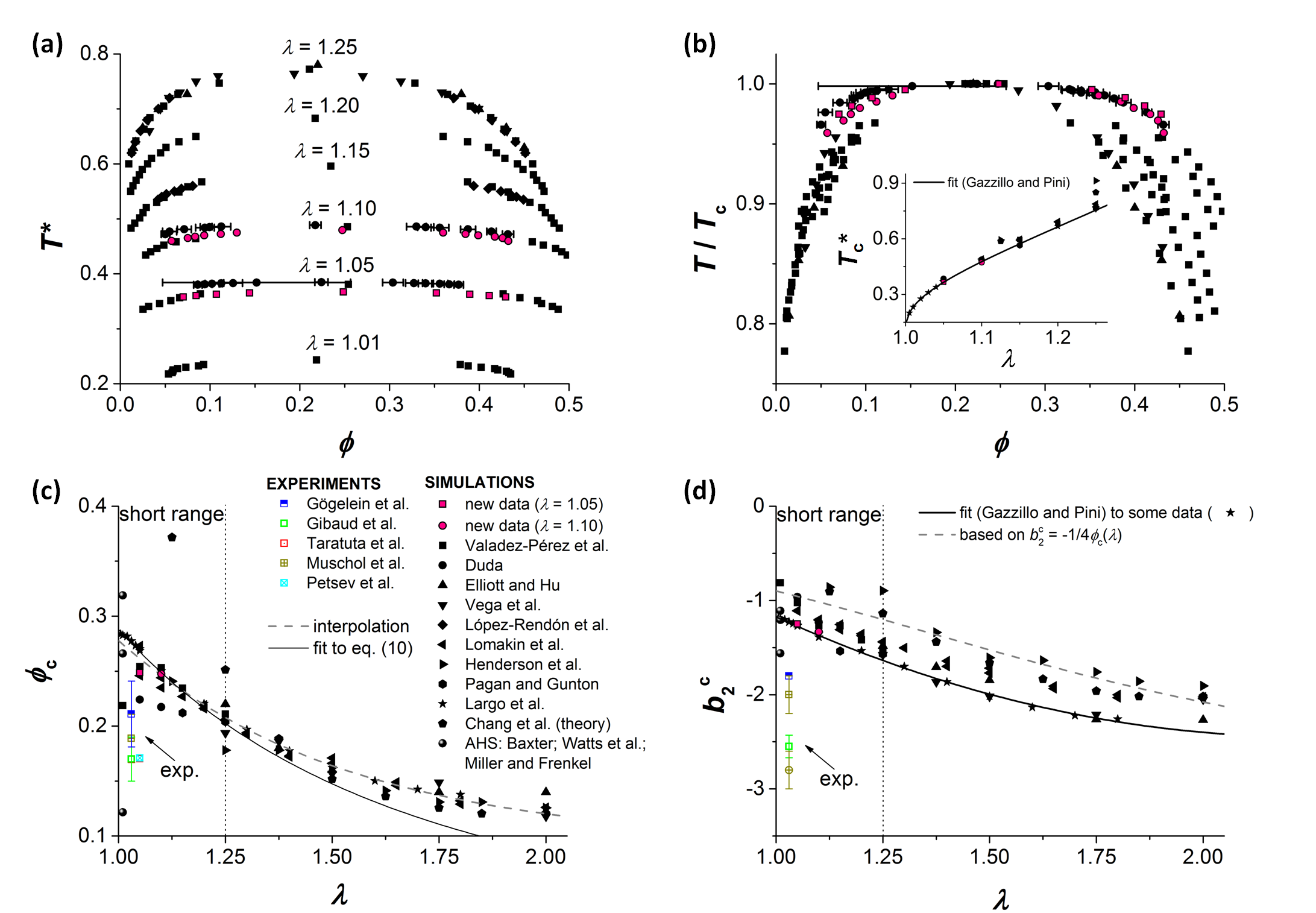}
  \caption{Square-well (SW) fluids with depth $\epsilon$ and range $\lambda$ (filled symbols). 
	(a) Gas--liquid binodals of short-range SW fluids, shown as reduced temperature $T^\star=k_\text{B}T/\epsilon$ vs. volume fraction $\phi$. 
	(b) Gas--liquid binodals of short-range SW fluids, re-plotted from (a), scaled to the critical temperature $T_\text{c}$. 
	(Inset) Reduced critical temperature $T^\star_\text{c}$ as a function of SW fluid range $\lambda$. 
	(c) Critical volume fraction $\phi_\text{c}$ of SW fluids as a function of their range $\lambda$. 
	(d) Normalized second virial coefficient of SW fluids at the critical point, $b_2^\text{c}$, as a function of their range $\lambda$. 
	In (c) and (d), open colored symbols represent values obtained in experiments on protein solutions (Tab.~\ref{tab:1}),\cite{Rosenbaum1999,Gibaud2011,Goegelein2012,Taratuta1990,Petsev2003} displayed at an arbitrary range. 
	New MC data (red symbols with black frames) and data taken from \cite{Valadez-Perez2012,Valadez-Perez2013,Duda2009,Elliott1999,Vega1992,Lopez-Rendon2006,Lomakin1996,Henderson1980,Pagan2005,Largo2008,Chang1994,Baxter1968,Watts1971,Miller2004,Gazzillo2013} as indicated.} 
  \label{fgr:1}
\end{figure}

In Figure~\ref{fgr:1}, we present new MC simulation data for $\lambda = 1.05$ and $1.1$ as well as simulation and perturbation theory data of square-well fluids collected from the literature.\cite{Valadez-Perez2012,Valadez-Perez2013,Duda2009,Elliott1999,Vega1992,Lopez-Rendon2006,Lomakin1996,Henderson1980,Pagan2005,Largo2008,Chang1994} 
Figure~\ref{fgr:1}a shows gas--liquid coexistence curves of short-range SW fluids as obtained by MC simulations, where the critical points have been estimated using  scaling laws and the law of rectilinear diameters.\cite{Panagiotopoulos1995}
It is important to note that data in the very vicinity of the critical point  ($0.95 \lesssim T/T_\text{c} \leq 1$) is quite scarce. 
Duda\cite{Duda2009} was able to perform $NVT-$MC simulations of SW fluids with $\lambda = 1.05$ and $1.10$ in this temperature window. 
Duda's values for the critical parameters are slightly off those obtained by Largo et al.\cite{Largo2008} 
Thus, we have done simulations using the Gibbs ensemble technique,\cite{Panagiotopoulos1987} as described previously.\cite{Valadez-Perez2012}
MC runs for SW fluids with $\lambda = 1.05$ and $1.1$ were performed (Tab.~\ref{tab:2}) in order to establish the SW binodal in the vicinity of $T_\text{c}$ and to provide estimates for the critical parameters, which agree with previous results\cite{Largo2008} (Fig.~\ref{fgr:1}c,d). 
The error bars of the new simulation data (red symbols with black frames) are smaller than the symbol size used in the plots, except the one for the critical points. 
With increasing range $\lambda$, the reduced coexistence temperature $T^\star = k_\text{B} T / \epsilon$ increases for all volume fractions studied due to enhanced attractions in fluids of larger range. 
This trend is also observed in the critical temperatures $T^\star_\text{c}$ (Fig.~\ref{fgr:1}b, inset).
Furthermore, a slight shift of the critical volume fraction $\phi_\text{c}$ to smaller values with increasing range $\lambda$ is observed (Fig.~\ref{fgr:1}c). 
The width of the coexistence region, limited by the gas (low $\phi$) and the liquid (high $\phi$) branches, slightly depends on $\lambda$. 

Figure~\ref{fgr:1}b shows the gas--liquid binodals of short-range SW fluids with temperatures normalized to the best estimates of the critical temperature.  
The gas branches of the different binodals seem almost to coincide within statistical error bars (in the low $\phi$ regime), although the curves are not scaled to the critical volume fraction $\phi_\text{c}$. 
However, the liquid branch of the coexistence curve moves toward smaller $\phi$ with increasing range $\lambda$ in this representation. 

Figure~\ref{fgr:1}c shows the best estimates of values of the volume fraction $\phi_\text{c}$ at the critical point (black symbols) as a function of interaction range $\lambda$, as determined by simulation and perturbation theory. 
We also include values reported for the AHS fluids, as obtained from the compressibility and energy route as well as simulation.\cite{Baxter1968,Watts1971,Miller2004} 
For a given range $\lambda$, a variety of values of $\phi_\text{c}$ ($\Delta \phi_\text{c} \approx 0.02-0.05$) have been reported for the SW fluid. 

It was recently proposed that the density at the critical point should be constant in the limit $\lambda \rightarrow 1$ if measured in units of the average distance between two bonded particles.\cite{Foffi2007,Largo2008} The simulation data of Figure~\ref{fgr:1}c have been scaled in this way (not shown). In the limit $\lambda \rightarrow 1$, we observe an approximately constant value
\begin{eqnarray}\label{eq:10}
\phi_\text{c}^0 = \phi_\text{c} \left( \frac{1+\lambda}{2} \right)^3 = 0.288,
\end{eqnarray}
which can be regarded as an approximation to the critical volume fraction of the AHS fluid. 
Our value for $\phi_\text{c}^0$ agrees with that reported by Largo et al.\cite{Largo2008} (0.289), but it is slightly larger than that for the AHS model predicted by simulations (0.266).\cite{Miller2003} 
For $\lambda \geq 1.5$, the values of $\phi_\text{c}$ are systematically larger than expected based on Eq.~(\ref{eq:10}) (Fig.~\ref{fgr:1}c, line).

For a given reduced coexistence temperature $T^\star$, the normalized $b_2$ value of a SW fluid of range $\lambda$ can be calculated analytically: 
\begin{eqnarray}\label{eq:sw}
b_2 (T^\star) = 1 - \left( \lambda^3 - 1 \right) \left(\exp{\left[ \frac{1}{T^\star} \right]} - 1 \right) .
\end{eqnarray}
Based on values for the reduced critical temperature $T_\text{c}^\star$ (Fig.~\ref{fgr:1}b, inset), we have calculated the second virial coefficient of SW fluids at the critical temperature $b_2^\text{c} = b_2 (T_\text{c}^\star)$ as a function of the range (Fig.~\ref{fgr:1}d, black symbols). 
The spread in $b_2^\text{c}$ for a given range $\lambda$ ($\Delta b_2^\text{c} \approx 1$) reflects different values reported for the critical temperatures $T_\text{c}^\star$ due to statistical and extrapolation errors as well as finite-size effects, which in general result in an overestimate of $T_\text{c}^\star$.\cite{Vega1992,Miller2004a} 

The recently proposed approximative relation\cite{Wolf2014} between $b_2^\text{c}$ and $\phi_\text{c}$, Eq.~(\ref{eq:1}), has been tested for the SW fluid. 
Based on the interpolation of simulation data of $\phi_\text{c}(\lambda)$ (Fig.~\ref{fgr:1}c, dashed line) and Eq.~(\ref{eq:1}), the $b_2^\text{c}$ values have been calculated (Fig.~\ref{fgr:1}d, dashed line). 
Despite the simplicity of the argument, the values obtained in this way are even in semi-quantitative agreement with simulation data of the SW fluid, although they are slightly too large. 
Similar agreement has been found for data of short-range HCAY and AO potentials (not shown). 
Furthermore, to fit the data of Largo et al.\cite{Largo2008} (Fig.~\ref{fgr:1}d, stars), Gazzillo and Pini\cite{Gazzillo2013} have used a second order polynomial (Fig.~\ref{fgr:1}d, solid line) 
\begin{eqnarray}\label{eq:3}
b_2^\text{c}(\lambda) = (b_2^\text{c})_0 + c_1 (\lambda - 1) + c_2 (\lambda - 1)^2 , 
\end{eqnarray}
where $(b_2^\text{c})_0$, $c_1$ and $c_2$ are fitting parameters to describe the range dependence of the $T^\star_\text{c}$ (Fig.~\ref{fgr:1}b, inset, line).

\begin{figure}
  	\centering
   \includegraphics[width=12cm]{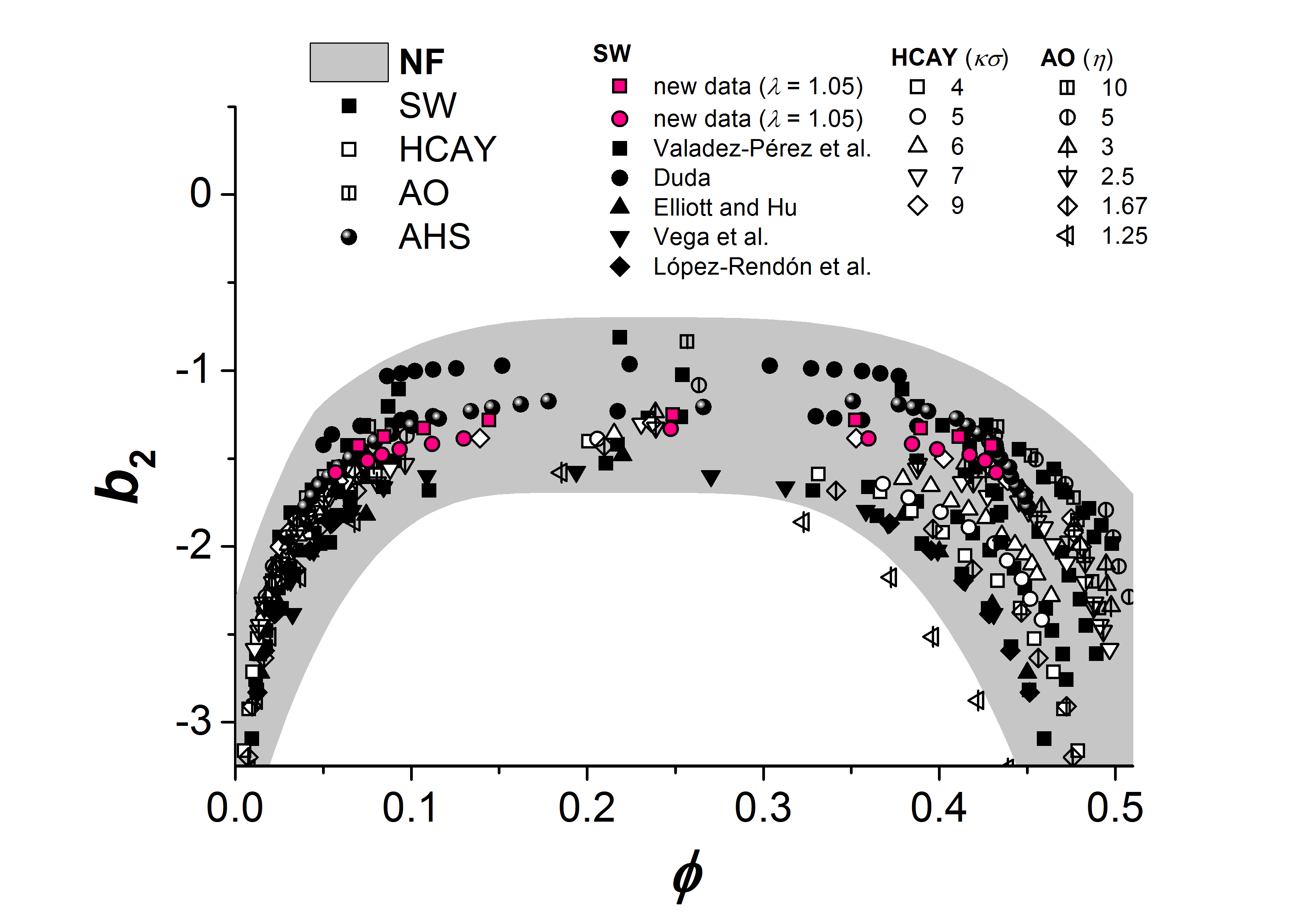}
  \caption{Binodals of short-range square-well (SW) fluids in the $b_2 - \phi$ plane for $1.01 \leq \lambda \leq 1.25$ (filled symbols, interpolated by a grey-shaded area as a guide to the eye).\cite{Valadez-Perez2012,Valadez-Perez2013,Duda2009,Elliott1999,Vega1992,Lopez-Rendon2006} 
	In addition, short-range hard-core attractive Yukawa (HCAY), Asakura-Oosawa (AO) and adhesive hard sphere (AHS) fluids are shown (open symbols, see legend).\cite{Valadez-Perez2012,Miller2004} 
	As suggested by the ELCS,\cite{Noro2000} the data for these potentials can be mapped onto short-range SW fluids, i.e., onto the grey shaded Noro-Frenkel (NF) area.} 
  \label{fgr:2}
\end{figure}

For the short-range SW fluids (Fig.~\ref{fgr:1}a), we have calculated binodals in the $b_2 - \phi$ plane, as shown in Figure~\ref{fgr:2}, i.e., we have replaced the $T^\star$ axis with the $b_2$ axis using Eq.~(\ref{eq:sw}). 
In addition to the SW potential, model potentials of liquid state theory, such as the hard-core attractive Yukawa (HCAY) potential\cite{Rosenbaum1996,Rosenbaum1999,Hagen1994,Valadez-Perez2012} and the Asakura-Oosawa (AO) potential,\cite{Lekkerkerker1992,Ilett1995,Kulkarni1999,Kulkarni2001} have also been used extensively to model interactions in protein solutions and colloidal suspensions. 
Here we focus on the gas--liquid phase separation only, although, for short-range potentials, crystallization intervenes and thus the gas--liquid coexistence becomes metastable.\cite{Dijkstra2002,Foffi2002,Tavares2004,Gast1983,Ilett1995,Lekkerkerker1992}
The HCAY potential can be written as follows
\begin{eqnarray}
U_\text{HCAY}(r) = \begin{cases}\infty, \quad & r < \sigma, \\-\epsilon \sigma \frac{\exp{[-\kappa(r-\sigma)]}}{r} , \quad & r \geq \sigma, \end{cases} 
\end{eqnarray}
where $\kappa$ quantifies the range of the HCAY potential. 
Sometimes, $(\kappa \sigma)^{-1}$ is used as an effective range of the HCAY model.\cite{Gazzillo2013} 
The main quantity characterizing the AO potential, which is often used in the context of colloid-polymer mixtures, is the size ratio $\eta$ between the colloid and the polymer (for details see, e.g., \cite{Ilett1995,Valadez-Perez2012}). 
Valadez-P\'erez et al.\cite{Valadez-Perez2012} have argued that short-range HCAY and AO potentials with $\kappa \sigma \gtrsim 3$ and $\eta \gtrsim 1.25$, respectively, correspond to  effective ranges $\lambda_\text{eff} \leq 1.25$, assuming a constant $b_2^\text{c}$ value of $-1.5$ (Eq.~(\ref{eq:vl})). 
For the effective range, we obtain, based on Eq.~(\ref{eq:3}), similar values (slightly larger ones for $\eta$, not shown). 

According to the Noro-Frenkel ELCS, it should be possible to map short-range HCAY and AO potentials onto short-range SW potentials, i.e., they should fall into the grey-shaded area of Figure~\ref{fgr:2}, which is a guide to the eye for the region covered by binodals of short-range SW fluids. 
For the parameter space of $\kappa$ and $\eta$ stated above, we have calculated binodals in the $b_2 - \phi$ plane (Fig.~\ref{fgr:2}) based on MC simulation data.\cite{Valadez-Perez2012} 
In addition, we include simulation results of the AHS potential.\cite{Miller2004} 
All these data, i.e. all gas and liquid branches of the binodals, lie in the grey-shaded region apart from one exception. 
(The liquid branch for the AO fluid with $\eta = 1.25$ lies slightly below the ELCS area; this value of $\eta$ seems to be too long-ranged.)
The data shown in Figure~\ref{fgr:2} provide further evidence for the mapping onto short-range SW fluids proposed by Noro and Frenkel.\cite{Noro2000} 
Moreover, the $b_2^\text{c}$ values fall -- by construction -- into a fairly narrow range around $-1.5$, which is in line with the VL suggestion (Eq.~(\ref{eq:vl})).\cite{Vliegenthart2000}

\section{Liquid--liquid phase separation: Gas--liquid binodals of protein solutions}\label{sec:3}
The generic phase diagram of an aqueous protein solution contains a solubility (liquidus) line, indicating -- for a given protein concentration $c_\text{p}$ -- the maximum temperature $T_\text{xtal}$ up to which protein crystals are stable or -- for a given temperature $T$ -- the maximum ``solubility'', i.e. the crystallization boundary. 
For $T > T_\text{xtal}$, a homogeneous protein solution without crystals represents the equilibrium state.\cite{Vekilov2012} 
(Note that thermal denaturation of proteins might interfere.) 
For temperatures below the crystallization boundary, metastable liquid--liquid phase separation (LLPS) of the protein solution into a protein-rich and a protein-poor phase is observed,\cite{Broide1996,Muschol1997} which is indicated by a dramatic clouding of the solution. 
At high protein concentrations, proteins might also form arrested gel or glassy states.\cite{Muschol1997,Sedgwick2005,Cardinaux2007,Gibaud2009,Gibaud2011} 
From a colloid physics point of view, proteins typically are regarded as colloids with short-range attractions and long-range electrostatic repulsion. 
In this respect, the LLPS coexistence curve corresponds to the gas--liquid binodal denoting the coexistence of a (dilute) colloidal gas and a (dense) colloidal liquid phase.\cite{Andersen2002} 

LLPS coexistence curves have been studied for proteins such as lysozyme, $\gamma$-crystallins and antibodies.\cite{Taratuta1990,Broide1996,Muschol1997,Thomson1987,Broide1991,Wang2013} 
In particular, lysozyme has often served as a model protein in physicochemical studies assessing the influence of different solution conditions, e.g., $p$H value, salt content or polymer concentration.\cite{Poon2000,Grigsby2001,Javid2007,Evers2008,Haehl2012,Kundu2015} 
Moreover, lysozyme is one of the most studied proteins with respect to modelling of protein--protein interactions.\cite{Curtis2002,Pagan2005,Goegelein2008,Duda2009,Goegelein2012,Valadez-Perez2012} 
Therefore, we will focus on the binodals of lysozyme solutions here, although our analysis can also be applied to other protein systems.

\begin{table}
\caption{\label{tab:1} Solution conditions under which the experimental gas--liquid binodals of lysozyme solutions have been determined: $p$H value, NaCl concentration $c_\text{s}$, buffer concentration $c_\text{b}$, additives (glycerol, DMSO, guandine hydrochloride). 
Temperatures at the critical point $T_\text{c}$ are obtained by fits based on Eq.~(\ref{eq:2}), typical error $\Delta T_\text{c} \approx 0.5~\text{K}$. Second virial coefficient at the critical point, $b_2^\text{c}$, if known from experiments.}
  	\centering
   \includegraphics[width=16.5cm]{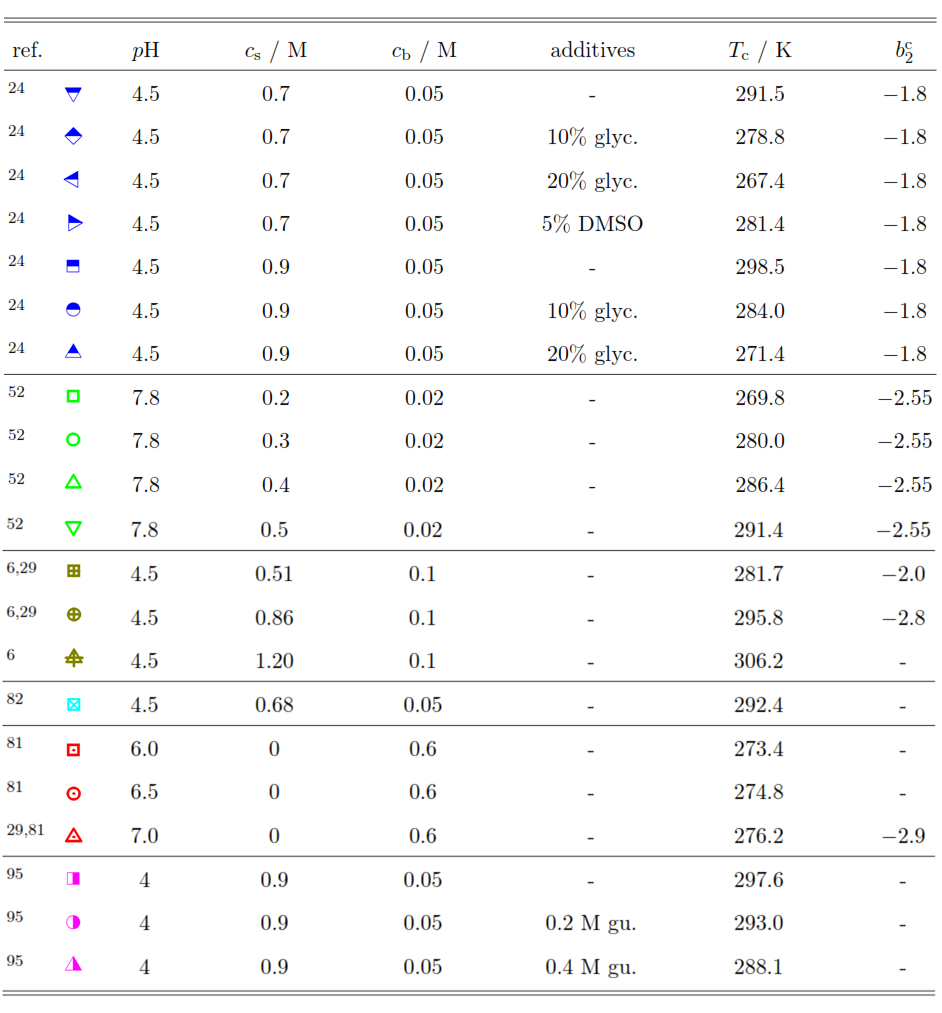}
\end{table}

We have collected a number of gas--liquid binodals of lysozyme solutions for various solutions conditions from the literature.\cite{Goegelein2012,Gibaud2011,Muschol1997,Petsev2003,Taratuta1990,Hansen2014}
In Table~\ref{tab:1}, we list solution conditions and interaction parameters of the experimental binodals, which we further analyze here. (We will use the values listed in Table~\ref{tab:1} for the DLVO model in Sec.~\ref{sec:6}.) 
The data sets cover a huge variety of different solutions conditions, i.e. $p$H values from 4 to 7.8, salt concentrations ranging from 0.2~M to 1.2~M and also include various additives, such as glycerol, DMSO and guanidine salt. 
This is important because tuning the solution conditions by salts and additives does not only affect protein--protein interactions, but also protein stability against aggregation and denaturation \cite{Chi2003,Shukla2011,Kamerzell2011,Liu2005,Evers2010,Voets2010,James2012} as well as protein adsorption;\cite{Koo2008,Evers2009a,Perriman2008,Huesecken2010,Evers2011} and therefore, insights into protein--protein interactions might in turn also be relevant for these aspects.

Published phase diagrams often give protein mass concentrations $c_\text{p}~[\text{mg}/\text{mL}]$ only. 
Hence, protein volume fractions $\phi$ have to be calculated. 
For all data considered here, we have calculated $\phi$ from $c_\text{p}$ assuming $\phi = c_\text{p} / \rho_\text{p}$, where $\rho_\text{p} = 1.351~\text{g}/\text{cm}^3$ is the protein mass density.\cite{Cardinaux2007,Gibaud2011,Goegelein2012} 
Note that slightly smaller values for $\rho_\text{p}$ are sometimes used in the literature, e.g. based on the molecular weight or different ways to quantify the volume of a lysozyme molecule in solution.\cite{Petsev2003,Sedgwick2005,Cardinaux2007}

\begin{figure}
  	\centering
   \includegraphics[width=18cm]{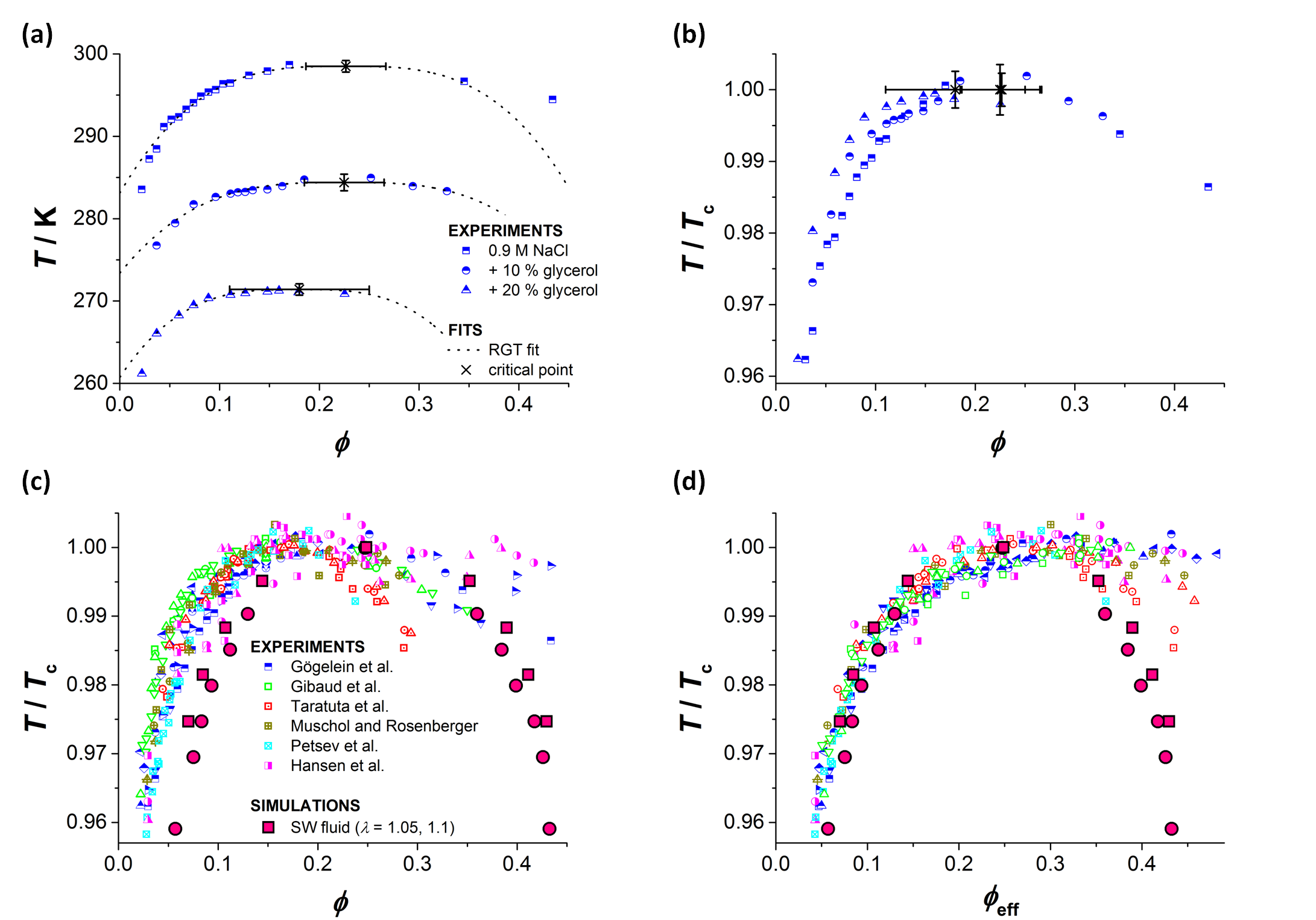}
  \caption{Experimental gas--liquid binodals of protein solutions (small symbols) compared to simulation data (large symbols). 
  (a) Representative experimental data,\cite{Goegelein2012} fitted by RGT scaling of Eq.~(\ref{eq:2}) (dashed lines) in order to estimate the critical points ($\phi_\text{c}, T_\text{c}$), which are indicated by crosses. 
  (b) Experimental data from (a), scaled to $T_\text{c}$. 
  (c) Experimental data for various solution conditions (Tab.~\ref{tab:1}) and simulation data for SW fluids with $\lambda = 1.05$ and $1.1$ (large symbols), scaled to $T_\text{c}$.  
  (d) Experimental binodals from (c) vs. effective volume fraction $\phi_\text{eff}$, scaled assuming a temperature-independent effective diameter $\sigma_\text{eff}$, and SW fluid binodals that do not require scaling. 
  Experimental data taken from \cite{Goegelein2012,Gibaud2011,Taratuta1990,Muschol1997,Petsev2003,Hansen2014}.} 
  \label{fgr:3}
\end{figure} 

To illustrate gas--liquid binodals of protein solutions on an absolute temperature scale, we use the binodals of lysozyme solutions in the presence of 0.9~M NaCl as well as various amounts (vol.\%) of glycerol (Fig.~\ref{fgr:3}a, symbols). 
In order to estimate the critical point from experimental data, binodals of protein solutions are often fitted by
\begin{eqnarray}\label{eq:2}
T(\phi) = T_\text{c} \left( 1 - \alpha \left| \frac{\phi-\phi_\text{c}}{\phi_\text{c}} \right|^{1/\beta} \right),
\end{eqnarray}
where $\alpha$ is a fitting parameter and $\beta = 0.325$ is the critical exponent for binary demixing from renormalization-group theory (RGT).\cite{Broide1996,Muschol1997} 
We note that this equation, based on critical phenomena, is valid only in the very proximity of the critical point, where the different shapes of the gas and liquid branches can be neglected. 
Previous experiments have suggested that $0.15 \lesssim \phi_\text{c} \lesssim 0.21$ (Fig.~\ref{fgr:1}c, colored symbols at an arbitrary short-range value of $\lambda$),\cite{Goegelein2012,Gibaud2011,Taratuta1990,Muschol1997,Petsev2003} which is slightly lower than those of the short-range SW fluids (black symbols: $0.2 \lesssim \phi_\text{c} \lesssim 0.28$).
To ensure that only data in the very vicinity of the critical point are included, the fitting procedure is restrained to experimental data in the region $0.05 \leq \phi \leq 0.35$. 
For data at smaller or larger values of $\phi$, the binodals show significant deviations from Eq.~(\ref{eq:2}) which would systematically distort the fitting.
The data shown in Figure~\ref{fgr:3}a have been fitted in this way (dashed lines) and the estimated critical points are indicated by crosses. 
A qualitative inspection reveals that the data around the critical point are well fitted, but indeed, at very low and very high volume fractions, we observe deviations between experimental binodals and RGT fits. 
This is due to the fact that Eq.~(\ref{eq:2}) assumes a certain width ($\beta$) and a symmetric binodal with respect to the critical point. 
While some studies reported symmetric binodals with respect to the critical point,\cite{Taratuta1990} a flattening of protein binodals in the liquid branch (at high volume fractions) can be observed in Figure~\ref{fgr:3} in agreement with other studies.\cite{Gibaud2011,Goegelein2012,Hansen2014}  

The RGT fits to cloud-point measurements (Fig.~\ref{fgr:3}a) yield values of the critical temperature $T_\text{c}$, which can be estimated with a typical uncertainty of $\Delta T_\text{c} \approx 0.5~\text{K}$ (Tab.~\ref{tab:1}).  
Therefore, the critical temperature $T_\text{c}$ turns out to be a reasonable scaling parameter.
In contrast, it is far more difficult to estimate reliable values of the critical volume fraction $\phi_\text{c}$, in parts due to the occasionally observed flattening of protein binodals. 
Typical uncertainties obtained from the fitting procedure are $\Delta \phi_\text{c} \approx 0.02-0.06$, i.e., up to $\sim 30~\%$ relative uncertainty. 
Therefore, a meaningful scaling of experimental data to the critical volume fraction can be difficult.
  
In Figure~\ref{fgr:3}b, we have scaled the data of Figure~\ref{fgr:3}a to the critical temperature $T_\text{c}$ and indicated the critical points as well as their estimated error bars. 
The different binodals have a similar shape, but their gas branches do not collapse, which is in contrast to the binodals of short-range SW fluids (Fig.~\ref{fgr:1}b).
We then have included further data sets with very different solution conditions (Tab.~\ref{tab:1}) and scaled their experimental binodals to their respective critical temperatures (Fig.~\ref{fgr:3}c).
Again, the scaled experimental data do not collapse onto a single master curve; at a given volume fraction $\phi > 0.05$, we rather observe a certain distribution of $T$ values ($\Delta T/T_\text{c} \approx 0.01$). 
A comparison with (simulations of) SW fluids is difficult, because most experimental data of protein binodals are limited to a very narrow temperature window ($0.95 \lesssim T/T_\text{c} \leq 1$), while simulation data in this temperature window, i.e. close to the critical point, is scarce. 
Hence, to allow for a direct comparison, we added our new MC simulation results for SW fluids with $\lambda = 1.05$ and $1.1$ for $T \lesssim T_\text{c}$ (Fig.~\ref{fgr:3}c, large symbols). 
The gas--liquid binodals of SW fluids are narrower than those of protein solutions. 

Different approaches to compare SW fluids with experiments on proteins have been followed in literature due to the variety of possible scaling parameters (e.g., $T_\text{c}, \phi_\text{c}, \epsilon, \lambda$).
Lomakin et al.\cite{Lomakin1996} aimed at modelling liquid--liquid phase separation of protein solutions. Matching the volume fraction at the critical point of the protein solution, they chose a SW fluid with $\lambda = 1.25$. However, the binodal of the corresponding SW fluid is much narrower than the experimentally determined one. 
They thus introduced a temperature dependent attraction strength $\epsilon (T) = \epsilon_\text{c} [ 1 + \gamma (T-T_\text{c})/T_\text{c} ]$ (with $\gamma = -3$) to ``broaden'' the coexistence curve of the SW fluid. 
We have transferred this approach to model the temperature dependence of experimentally determined second virial coefficients of protein solutions (Sec.~\ref{sec:4}, Fig.~\ref{fgr:4}b, dotted line).
However, this procedure works only if the $b_2^\text{c}$ values of experiments and simulations agree with each other (cf. Fig.~\ref{fgr:1}d). 
Grigsby et al.\cite{Grigsby2001} used cloud-point temperature (CPT) measurements and a perturbation theory approach in order to estimate how the strength of the attraction $\epsilon$ of a SW fluid with $\lambda = 1.2$ varies for different salts as a function of their concentration. 
Wentzel and Gunton\cite{Wentzel2007a} scaled experimental protein and simulation binodals ($\lambda = 1.15$) at a specific number density and used CPT measurements at this density to calculate the liquid--liquid coexistence surface in protein--salt solutions and to determine the salt-dependence of the second virial coefficient. 
However, the predicted coexistence surfaces seem to be too narrow compared with experimental data (as in Fig.~\ref{fgr:3}c) and this model implicitly assumes that the temperature dependence of experimental $b_2$ values is well described by the corresponding relation of SW fluids (cf. Fig.~\ref{fgr:4}b). 
As suggested by the van der Waals corresponding states law, often binodals of SW fluids and protein solutions are not only scaled to the critical temperature, but also to the critical volume fraction,\cite{Pagan2005,Duda2009,Valadez-Perez2012,Liu2007} although the value of $\phi_\text{c}$ can only be determined with a large uncertainty in many cases.
Nevertheless, Duda\cite{Duda2009} could show that the binodal of very short-range SW fluids ($\lambda = 1.05$ and $1.10$) nicely agrees with experimentally determined binodals of lysozyme and $\gamma$-crystallins (and also with a binodal of a patchy fluid\cite{Liu2007}), while binodals of SW fluids with $\lambda = 1.15$ are too narrow and only qualitatively capture features of the protein binodals.\cite{Pagan2005,Liu2007} 

A qualitative difference between the SW potential and the protein--protein interaction potential is the presence of a repulsive contribution to the latter.
According to the Noro-Frenkel ELCS,\cite{Noro2000} it should be possible to map the binodals of protein solution onto effective SW fluids if the repulsive contribution of the interaction potential is taken into account in terms of an effective hard-core diameter $\sigma_\text{eff}$, as defined in Eq.~(\ref{eq:eff}). 
However, the protein--protein interaction potential $U(r)$ is usually not known and therefore this procedure, i.e., to calculate $\sigma_\text{eff}$ from $U(r)$, has not been applied to experimental data so far.
Here, we determine $\sigma_\text{eff}$ by a comparison of experimental binodals with our simulation data of the SW fluid with $\lambda = 1.05$. 
We fit each gas branch of the protein binodals (Fig.~\ref{fgr:3}c) to the simulation data using only one fit parameter, namely $\sigma_\text{eff}$.  
This implies that $\sigma_\text{eff}$ is assumed to be temperature-independent, although, in general, the effective hard-core diameter might be temperature dependent. 
(In Sec.~\ref{sec:6} (Fig.~\ref{fgr:7}a), we compare the values of $\sigma_\text{eff}$ with calculations of $\sigma_\text{eff}^\text{DLVO}(T)$ based on DLVO theory.) 
The parameter $\sigma_\text{eff}$ might be related to the fitting parameter $\alpha$ of Eq.~(\ref{eq:2}), describing the width of the binodal. 
However, as the value of $\alpha$ is strongly correlated with the value of $\phi_c$, which is retrieved by fitting, $\alpha$ is not known with high accuracy. 

Figure~\ref{fgr:3}d shows the scaled protein binodals as a function of effective volume fraction $\phi_\text{eff} = ( \sigma_\text{eff} / \sigma )^3 \phi$. 
All experimental data fall on a single master curve in the gas branch, which -- by construction -- is given by the binodal of a short-range SW fluid. 
Since the binodals of the short-range SW fluids almost collapse in the low density part (Fig.~\ref{fgr:1}b), this is, to a good approximation, valid for all short-range SW fluids with $\lambda \lesssim 1.25$.
Interestingly, the assumption of a temperature-independent hard-core diameter seems to work quite well. 
We note that the experimental data in the liquid branch show a wide spread. 
While, for some data sets (blue, green and magenta symbols), we observe $T/T_\text{c} \geq 0.995$ even for $\phi > 0.4$, which corresponds to a significant flattening of the liquid branch, other data sets (red and dark yellow symbols) show roughly symmetric gas--liquid binodals almost agreeing with the liquid branch of the SW fluid. 
The differences of the experimental binodals at high volume fractions are mainly ascribed to experimental difficulties in the determination of CPT at large $\phi$.\cite{Cardinaux2007}

To test whether the experimental data obey the Noro-Frenkel predictions, they need to be plotted as a function of second virial coefficient $b_2$ instead of temperature $T$ (Sec.~\ref{sec:5}). Thus, we first discuss the temperature dependence of $b_2$ (Sec.~\ref{sec:4}).


\section{Temperature dependence of the second virial coefficient of protein solutions}\label{sec:4}
\begin{figure}
  	\centering
   \includegraphics[width=18cm]{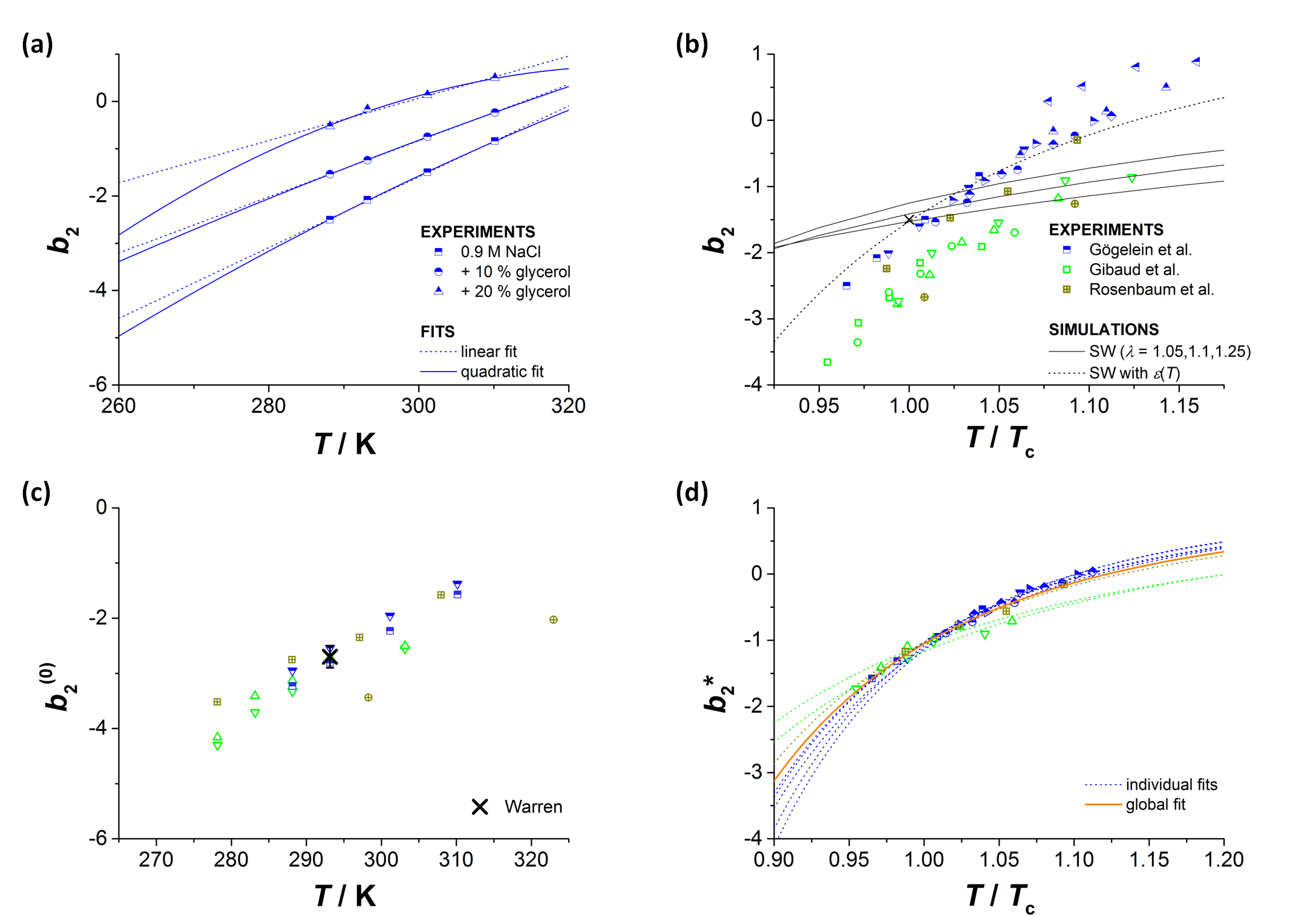}
  \caption{Temperature dependence of the second virial coefficient $b_2$ of protein solutions. 
  (a) Representative experimental data (symbols, solution conditions as in Fig.~\ref{fgr:3}a).\cite{Goegelein2012} Heuristic linear and quadratic fits (dotted and solid lines). 
  (b) Experimental data for various solution conditions\cite{Goegelein2012,Gibaud2011,Rosenbaum1999} (symbols) scaled to the critical temperature $T_\text{c}$ of the corresponding gas--liquid binodal (Tab.~\ref{tab:1}, Fig.~\ref{fgr:3}c), data for SW fluids with $\lambda = 1.05, 1.1, 1.25$ (lines, top to bottom) and $\lambda = 1.25$ with $\epsilon(T)$ (dotted line).\cite{Lomakin1996} 
  (c) Second virial coefficient of the baseline model, $b_2^{(0)}$, as obtained from the Donnan scaling (Eq.~(\ref{eq:7})). 
  (d) Reduced second virial coefficient $b_2^\star$ that accounts for effective volume fraction $\phi_\text{eff}$ (from Fig.~\ref{fgr:3}d). Fits to Eq.~(\ref{eq:6}) assuming SW fluids with a $T$ dependent depth (lines).} 
  \label{fgr:4}
\end{figure}

In order to present gas--liquid binodals in the $b_2 - \phi$ plane, one has to replace the temperature $T$ with the second virial coefficient $b_2$, i.e., the relation $b_2(T)$ between temperature and second virial coefficient has to be known at all cloud-point temperatures. 
In practice, the second virial coefficient is measured at a couple of temperatures only, because the determination of $b_2$ values, e.g. by static light scattering, represents an experimental challenge due to the small size of proteins and the drastic effects of impurities or aggregates on the scattering intensity. 
Typical experimental uncertainties of $b_2$ values are on the order of $\Delta b_2 \approx 0.2-0.4$.\cite{Rosenbaum1999} 
In Figure~\ref{fgr:4}a, we show $b_2(T)$ measurements corresponding to the solution conditions of the binodals of Figure~\ref{fgr:3}a as a typical example.\cite{Goegelein2012} 
In many cases, an approximate relation $b_2(T)$ is established based on an interpolation or even an extrapolation of the measured $b_2$ values. 
As the functional form of $b_2(T)$ is not known a priori, heuristic fitting of linear\cite{Rosenbaum1996,Rosenbaum1999} or quadratic functions\cite{Goegelein2012} to $b_2(T)$ data has been proposed. 
Moreover, it has been proposed that the stickiness parameter $\tau = 1/[4(1-b_2)]$ of protein solutions follows a linear temperature dependence.\cite{Piazza1998} 
We use linear and quadratic functions to fit the data shown in Figure~\ref{fgr:4}a, illustrating the effects of different proposed $b_2(T)$ relations. 
Depending on the fitting procedure, systematically different $b_2$ values might be predicted, in particular, if an extrapolation of the measured data to higher and especially lower temperatures is needed. 
The discrepancies between heuristic fitting approaches call for a theoretical reasoning of the temperature dependence of $b_2$. 

In Sec.~\ref{sec:3}, we have shown that protein binodals lie near a master curve if the temperature axis is scaled to the critical temperature $T_\text{c}$ (Fig.~\ref{fgr:3}c,d). 
Therefore, one might speculate that the $b_2(T)$ curves (or $b_2^\star(T)$ curves) for different solution conditions lie near a master curve when the temperature axis is scaled to $T_\text{c}$, in particular for $T < T_\text{c}$, such that the ELCS mapping on equivalent SW fluids is fulfilled. 
In this case, experimental data scaled to the $b_2^\star - \phi_\text{eff}$ plane would fall onto the (grey-shaded) area of SW fluids in Figure~\ref{fgr:2}.

We have collected $b_2(T)$ curves for a number of solution conditions in the literature,\cite{Goegelein2012,Gibaud2011,Rosenbaum1999} for which also the critical temperature of the gas--liquid binodal is known\cite{Goegelein2012,Gibaud2011,Muschol1997} and at least two $b_2$ values (per solution condition) in the vicinity of the critical point ($0.95 \leq T / T_\text{c} \leq 1.05$) are available in order to avoid extrapolation errors.
Moreover, we focus on data in the high-salt regime ($c_\text{s} > 0.3~\text{M}$),\cite{Vekilov2012} where the range of electrostatic interactions is at most about 20~\% of the particle size and hence the interaction potential sufficiently short-range such that the ELCS applies.
It is important to note that the experimental values of $b_2$ at the critical temperature, $-1.7 \leq b_2^\text{c} \leq -2.9$,\cite{Rosenbaum1999,Gibaud2011,Goegelein2012} are systematically lower than the values of short-range SW fluids, $-0.8 \leq b_2^\text{c} \leq -1.6$ (Fig.~\ref{fgr:1}d).
(We note that Rosenbaum et al.\cite{Rosenbaum1999} have also estimated $b_2^\text{c}$ values of $-0.8$ and $-1.2$ for specific solution conditions.)
Nevertheless, we have scaled the collected $b_2(T)$ curves to the respective critical temperatures (Fig.~\ref{fgr:4}b). 

The data of G\"ogelein et al.\cite{Goegelein2012} (with the exception of one data set measured only at high $T/T_\text{c}$ values) and those of Gibaud et al.\cite{Gibaud2011} collapse on two different master curves (Fig.~\ref{fgr:4}b, blue and green symbols, respectively). 
The two data sets are based on very different salt concentrations and $p$H values. 
To take these differences into account, we recall the Donnan argument, which has been used to explain salt and charge scaling of protein interactions.\cite{Poon2000,Warren2002} 
In particular, a scaling collapse of the crystallization boundary and the second virial coefficients of lysozyme solutions at room temperature has been observed.\cite{Poon2000,Warren2002} 
Warren\cite{Warren2002} has proposed a theoretical background for the charge-salt concentration scaling of protein interactions. 
He particularly assumed that the free energy of a protein solution is given by a baseline model, e.g., the AHS model or Sear's sticky particle model,\cite{Sear1999} plus terms depending on the charge and concentration of proteins and the salt concentration. 
In this way, the second virial coefficient in the high-salt limit is given by
\begin{eqnarray}\label{eq:7}
B_2 = B_2^{(0)} + \frac{Q^2}{4 c_\text{s}} ,
\end{eqnarray}
where $B_2^{(0)}$ is the second virial coefficient of the baseline model and $Qe$ the charge of the protein with $e$ the electronic charge, which is known from titration experiments.\cite{Tanford1972,Kuehner1999}  
At room temperature (and high salt concentration), Warren found $b_2^{(0)} = B_2^{(0)} / B_2^{\text{HS}} \approx -2.7 \pm 0.2$. 
A similar scaling is found for computer simulations of the so-called primitive model.\cite{Allahyarov2003} 
As the argument for Eq.~(\ref{eq:7}) also holds at another temperature, we expect another $b_2^{(0)}$ value and thus we assume to find an almost universal $b_2^{(0)}(T)$ curve. 
Based on the specific solution conditions, namely $c_\text{s}$ and the $p$H (or $Q$\cite{Tanford1972,Kuehner1999}), we have calculated $b_2^{(0)}$ values for the data shown in Figure~\ref{fgr:4}b. 
Indeed, in Figure~\ref{fgr:4}c, we observe that, at a given temperature $T$, data from different solution conditions approximately take the same $b_2^{(0)}$ value, which is -- at room temperature -- in agreement with the value determined by Warren. 
(For clarity, we do not include further literature data\cite{Curtis1998,Bonnete1999,Tessier2002,Parmar2009} in this figure, which would lead to an even larger spread of the data; the spread of $b_2^{(0)}$ is ascribed to the crude assumption of a quasi-universal baseline model and to experimental difficulties in determining $b_2$ values.) 
The two different master curves observed in Figure~\ref{fgr:4}b (of G\"ogelein et al.\cite{Goegelein2012} and Gibaud et al.\cite{Gibaud2011}) are thus due to the very different solution conditions and follow roughly the Donnan scaling.

The solid lines in Figure~\ref{fgr:4}b show the temperature dependence of SW fluids with $\lambda = 1.05, 1.1, 1.25$ (top to bottom). 
The temperature dependence of $b_2$ for the short-range SW fluids is much weaker than that of the protein solutions. 
Nevertheless, one could relate the experimental $b_2(T)$ curves to SW fluids by introducing a temperature dependent range $\lambda(T)$, as suggested by Noro and Frenkel.\cite{Noro2000} 
It has also been proposed to introduce a linear temperature dependence of the SW depth $\epsilon(T)$ (Fig.~\ref{fgr:4}b, dotted line),\cite{Lomakin1996} which was motivated to compensate for the too narrow width of SW binodals (compared to those of protein solutions).

The interactions in SW fluids can account for attractions only, while protein interactions comprise attractive and repulsive contributions.
To take repulsive protein--protein interactions into account, Noro and Frenkel\cite{Noro2000} have proposed to introduce an effective hard-core diameter $\sigma_\text{eff}$. 
We have determined values of $\sigma_\text{eff}$ for all protein binodals in Sec.~\ref{sec:3}. 
For the data shown in Figure~\ref{fgr:4}b, we compute the repulsion-corrected $b_2$, i.e. $b_2^\star$, based on Eq. (\ref{eq:4}); the results are shown in Figure~\ref{fgr:4}d. 
(Note that the $b_2^\star$ values slightly depend on the value of the protein mass density $\rho_\text{p}$ used to calculate $\phi$; a smaller $\rho_\text{p}$ leads to slightly smaller $b_2^\star$.)
All data sets in Figure~\ref{fgr:4}d, independent of their solution conditions, almost follow a master curve in the $b_2^\star - T/T_\text{c}$ representation, although the green data sets\cite{Gibaud2011} show slightly higher $b_2^\star$ values for $T < T_\text{c}$ and lower one for $T > T_\text{c}$ than the blue ones.\cite{Goegelein2012} 
Since $b_2^\star$ takes all repulsive contributions into account, this includes the electrostatic effect discussed above (Eq.~(\ref{eq:7})).

We then propose to model the $b_2^\star (T / T_\text{c})$ curve using the temperature dependence of a SW fluid (Eq.~(\ref{eq:sw})) assuming a linear $T$ dependence of the SW depth:
\begin{eqnarray}\label{eq:6}
b_2^\star \left( \frac{T}{T_\text{c}} \right) = 1 - (\lambda^3 - 1) \left( \exp{\left[ \frac{a T / T_\text{c} + b}{T / T_\text{c}} \right]} - 1 \right) ,
\end{eqnarray}
where we set $\lambda = 1.05$ and use $a$ and $b$ as fitting parameters.
Assuming that all data points fall on a single curve, we fit this expression to all data (Fig.~\ref{fgr:4}d, solid orange line) and obtain $a = -3.3$ and $b = 5.9$. 
If it is fitted to the data sets individually, the results (dashed lines) show only slightly better agreement with the data. 

It has been argued that (experimental) $b_2$ values of protein solutions might only show a weak temperature dependence, mainly governed by the Boltzmann factor, and that other temperature effects, e.g. of hydration, are of minor importance.\cite{Piazza1998,Malfois1996} 
Despite the presumably weak temperature dependence of $b_2$, it is conceivable that, for $T \gg T_\text{c}$, significant deviations from the master curve (Fig.~\ref{fgr:4}d, orange line) occur for other solution conditions. 
Since our analysis contains a limited number of data sets only, this remains to be tested in future experiments.


\section{The extended law of corresponding states for protein solutions}\label{sec:5}
In Sec.~\ref{sec:3}, we have shown that the gas--liquid binodals of protein solutions scaled to the critical temperature $T_\text{c}$ fall on a single master curve if a temperature independent effective hard-core diameter $\sigma_\text{eff}$ is introduced (Fig.~\ref{fgr:3}d). 
In Sec.~\ref{sec:4}, we have argued that the temperature dependence of the second virial coefficient $b_2(T)$ or $b_2^\star(T)$ can be explained based on a SW fluid with a $T$ dependent depth. In view of this analysis, we are now able to present the experimental binodals of protein solutions in the $b_2^\star - \phi_\text{eff}$ plane. 
   
\begin{figure}
  	\centering
   \includegraphics[width=12cm]{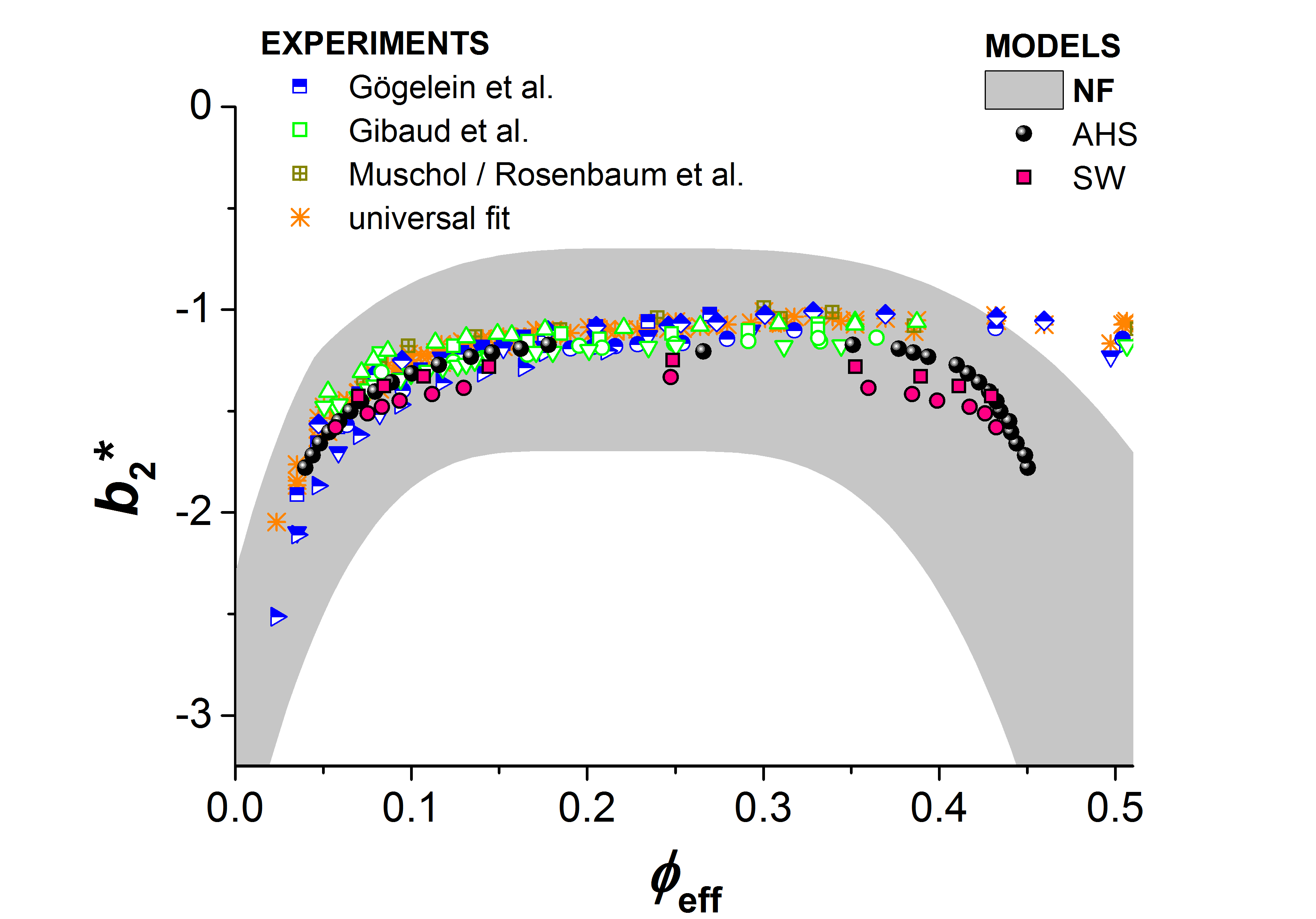}
  \caption{Gas--liquid binodals, shown as $b_2^\star$ vs. $\phi_\text{eff}$: comparison of experimental data (open colored symbols) and simulation results of the adhesive hard-sphere (AHS) fluid (black symbols) and short-range square-well (SW) fluids (red symbols with black frames) with the grey shaded Noro-Frenkel (NF) area. Data taken from \cite{Goegelein2012,Gibaud2011,Muschol1997,Rosenbaum1999,Miller2004}. } 
  \label{fgr:5}
\end{figure}

Based on the master curve for the dependence of $b_2^\star$ on $T$ (Fig.~\ref{fgr:4}d, orange line), we obtain a master curve of experimental protein binodals in the $b_2^\star - \phi$ plane (Fig.~\ref{fgr:5}, orange stars).  
If one allows for slight variations of the $T$ dependence of $b_2^\star$ for different solution conditions (as implied by the individual fits in Fig.~\ref{fgr:4}d), each binodal is scaled separately (colored symbols), including data sets with low salt content. 
The data based on individual fits of $b_2^\star(T)$ almost collapse onto the universal curve (orange stars), indicating that the universal $b_2^\star(T)$ curve indeed provides a reasonable description of the data in the relevant temperature window. 
In addition to the experimental data in the $b_2^\star - \phi_\text{eff}$ representation, we re-plot the Noro-Frenkel area of short-range SW fluids (from Fig.~\ref{fgr:2}) as well as the simulation binodals of the AHS fluid and our SW fluid data ($\lambda = 1.05,1.1$) on the same axes as in Figure~\ref{fgr:2} (taking into account that, in these cases, $\phi_\text{eff} = \phi$).
The experimental protein data are completely encompassed by the (grey-shaded) Noro-Frenkel area of short-range SW fluids.
It is striking that the experimental data seems to follow very closely the binodal of the AHS model (except for very high volume fractions $\phi_\text{eff} \gtrsim 0.4$). 
The deviations at high volume fractions might be related to the appearance of non-equilibrium states.\cite{Sedgwick2005,Gibaud2009,Vekilov2012}

Figure~\ref{fgr:5} represents the main result of this work. 
For a broad set of experimental conditions, it shows that, based on a one-parameter scaling of experimental protein data to short-range SW fluids, which yields $\sigma_\text{eff}$ (Sec.~\ref{sec:3}), and an adequate description of the temperature dependence of the second virial coefficient, $b_2^\star (T / T_\text{c})$ (Sec.~\ref{sec:4}), protein phase behavior appears to quantitatively agree with the predictions of the ELCS, as proposed by NF.\cite{Noro2000} 
Thus, the binodals of protein solutions can be mapped onto those of short-range SW fluids if repulsive particle--particle interactions are taken into account in terms of an effective hard-core diameter $\sigma_\text{eff}$. 
Moreover, the proposed way of quantifying repulsive protein--protein interactions is able to clarify the discrepancies between short-range SW fluids and experimental protein data (Sec.~\ref{sec:2} and Fig.~\ref{fgr:1}c,d). 

\section{DLVO model of protein--protein interactions}\label{sec:6}
DLVO theory is frequently used to model protein--protein interactions.\cite{Muschol1995,Farnum1999,Sedgwick2007,Goegelein2012} The DLVO approach is more complex than SW fluid modelling, but DLVO theory contains attractive van der Waals as well as repulsive electrostatic interactions, which appears to be more realistic than simple SW fluid models.  

The DLVO interaction potential $U_\text{DLVO}(r)$\cite{Goegelein2012} is defined as
\begin{eqnarray}
U_\text{DLVO}(r) = U_\text{HS}(r) + U_\text{SC}(r) + U_\text{VDW}(r) .
\end{eqnarray}
The hard-sphere contribution $U_\text{HS}(r)$ is given by
\begin{eqnarray}
U_\text{HS}(r) = \begin{cases}\infty, \quad & r < \sigma, \\0, \quad & r \geq \sigma.\end{cases} 
\end{eqnarray}
For $r > \sigma$, the screened Coulomb contribution $U_\text{SC}(r)$ is given by
\begin{eqnarray}
U_\text{SC}(r) = \frac{\left( Q e \right)^2}{4 \pi \epsilon_0 \epsilon_\text{s} r} \frac{\exp{\left[ -\kappa (r - \sigma) \right]}}{\left( 1 + \kappa \sigma/2 \right)^2} 
\end{eqnarray}
and the van der Waals contribution $U_\text{VDW}(r)$ by 
\begin{eqnarray}
U_\text{VDW}(r) = - \frac{A}{12} \left( \frac{\sigma^2}{r^2-\sigma^2} + \frac{\sigma^2}{r^2} + 2 \ln{\left[ 1 - \frac{\sigma^2}{r^2} \right]} \right),
\end{eqnarray}
where $\epsilon_0$ and $\epsilon_\text{s}$ are the permittivity of vacuum and the solvent, respectively.  
The permittivity of water has a weak temperature dependence, which we describe by $\epsilon_s = \epsilon_{\text{s},0} - \epsilon_{\text{s},1} \cdot (T-273.15~\text{K}) / \text{K}$ with $\epsilon_{\text{s},0}=87.62$ and $\epsilon_{\text{s},1}=0.36$ for $270~\text{K} \lesssim T \lesssim 330~\text{K}$, in agreement with previous work.\cite{Goegelein2012} 
For water--glycerol and water--DMSO mixtures, we obtained similar relations.\cite{Goegelein2012}
The Debye screening length $\kappa^{-1}$ is -- for monovalent salts -- given by 
\begin{eqnarray}\label{eq:dlvo}
\kappa^2 =  \frac{1}{1-\phi} \frac{e^2 N_\text{A}}{\epsilon_\text{s} \epsilon_0 k_\text{B} T} \left( \frac{Q \phi \rho_\text{p}}{M_\text{w}} + 2 c_\text{s} + 2 c_\text{b}   \right),
\end{eqnarray}
where $c_\text{s}$ and $c_\text{b}$ are the molar concentrations of salt and (dissociated) buffer, $N_\text{A}$ is Avogadro's number, and $M_\text{w}$ the molecular weight of the protein.
The strength of the van der Waals contribution is given by the Hamaker constant $A$, which for lysozyme interactions in brine is set to $A = 8.3~k_\text{B}T$ in agreement with previous studies.\cite{Poon2000,Sedgwick2007}

It has been shown that DLVO-based models are able to quantitatively account for protein phase behavior\cite{Pellicane2004a,Pellicane2012,Goegelein2012} if suitable values for the Hamaker constant are chosen. 
In previous work,\cite{Sedgwick2007} it has been shown that DLVO-based calculations of $b_2$ values are in very good agreement with experimental data\cite{Rosenbaum1999,Piazza2000a,Rosenbaum1996,Bonnete1999,Velev1998,Gripon1997} for $p$H~4.5 at room temperature. 
We now compare DLVO calculations with further experimental data and include data at $p$H~7.8. 

The second virial coefficient of the DLVO potential, $b_2^\text{DLVO}$, can be computed directly by  
\begin{eqnarray}\label{eq:dd}
b_2^\text{DLVO} = 1 + \frac{3}{\sigma^3} \int_{\sigma+\delta}^{\infty} \left( 1 - \exp{\left[- \frac{U_\text{SC}(r) + U_\text{VDW}(r)}{k_\text{B} T} \right]} \right) r^2 \text{d}r,
\end{eqnarray}
where $\delta$ represents a cut-off length which is introduced to avoid divergence of the integral. For consistency with previous work,\cite{Poon2000,Sedgwick2007} we use $\delta = 0.1437~\text{nm}$ at room temperature $T = 293.15~\text{K}$. 

\begin{figure}
  	\centering
   \includegraphics[width=12cm]{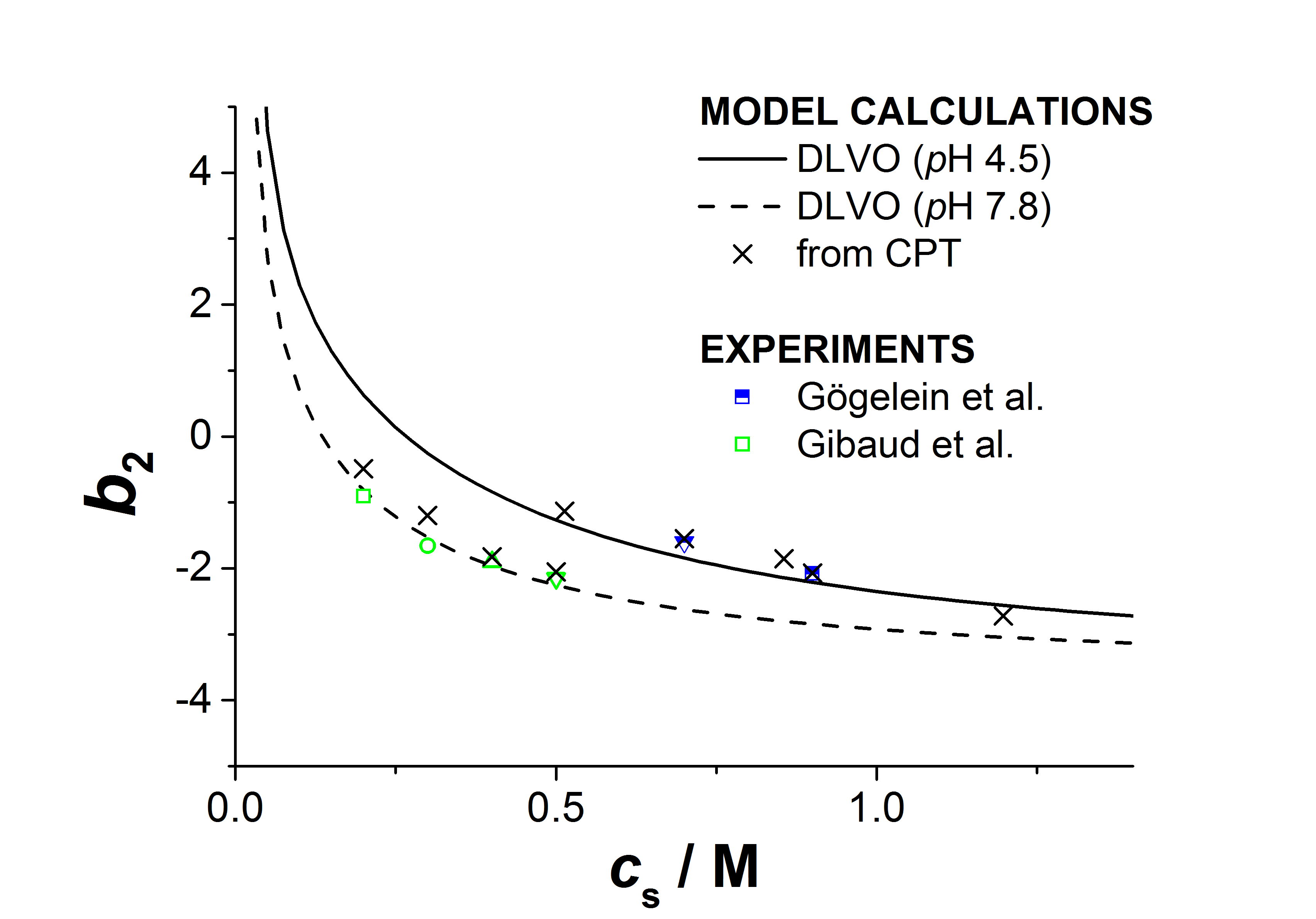}
  \caption{Second virial coefficient $b_2$ as a function of salt (NaCl) concentration $c_\text{s}$ at room temperature: determined by scattering experiments (open symbols),\cite{Gibaud2011,Goegelein2012} DLVO calculations based on Eq.~(\ref{eq:dd}) (lines), and from cloud-point temperature (CPT) measurements (crosses, Sec.~\ref{sec:7}).}
  \label{fgr:6}
\end{figure}

In Figure~\ref{fgr:6}, we compare experimental $b_2$ values, discussed in this work (open symbols),\cite{Gibaud2011,Goegelein2012} with calculations (lines) based on DLVO theory (Eq.~(\ref{eq:dd})). 
We find almost quantitative agreement between experiments and theory for both $p$H values and all salt concentrations studied. 
For clarity, we do not re-plot the experimental data,\cite{Rosenbaum1999,Piazza2000a,Rosenbaum1996,Bonnete1999,Velev1998,Gripon1997} which have been shown to closely follow the DLVO calculations at $p$H~4.5.\cite{Sedgwick2007}

\begin{figure}
  	\centering
   \includegraphics[width=18cm]{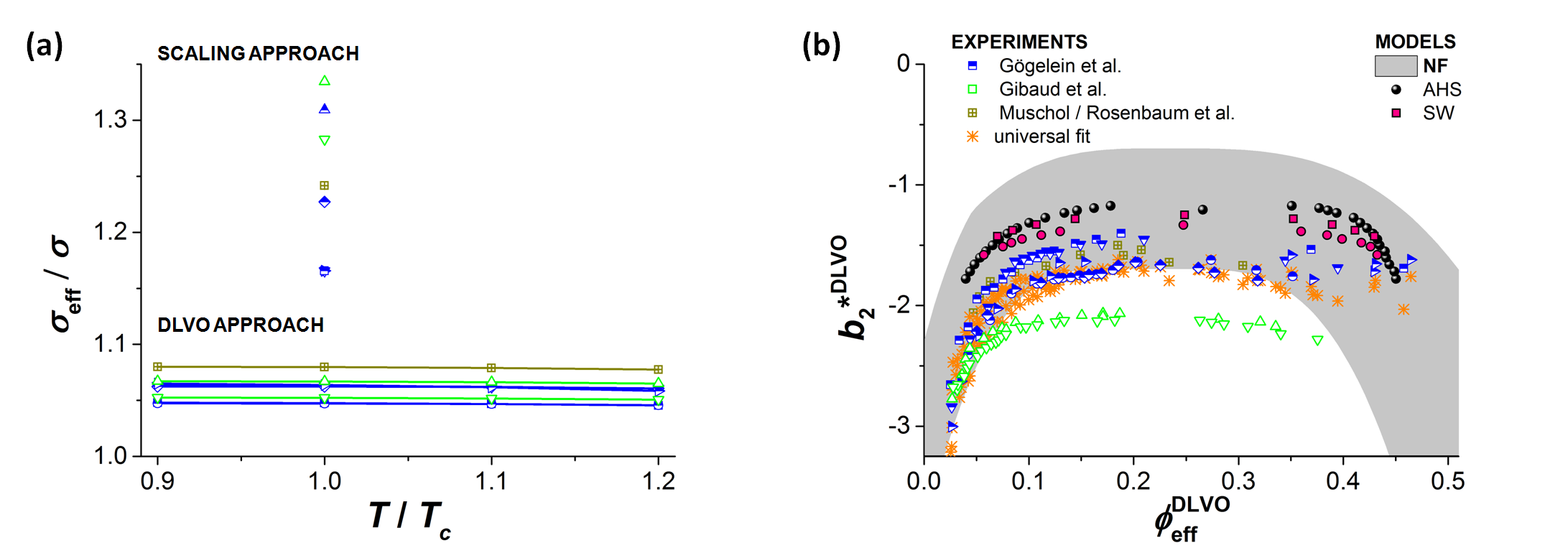}
  \caption{(a) Effective hard-core diameter $\sigma_\text{eff}$, taking into account repulsive interactions, as a function of temperature $T$: temperature-independent values determined by scaling approach (symbols, at an arbitrary temperature) and values determined assuming DLVO interactions (lines). 
	(b) Gas--liquid binodals, shown as $b_2^{\star\text{DLVO}}$ vs. $\phi_\text{eff}^\text{DLVO}$: comparison of experimental data (open colored symbols) with simulation results of AHS fluids (black symbols) and short-range SW fluids (red symbols with black frames) with the grey shaded Noro-Frenkel (NF) area. Data taken from \cite{Goegelein2012,Gibaud2011,Muschol1997,Rosenbaum1999,Miller2004}.} 
  \label{fgr:7}
\end{figure}

As DLVO theory is able to quantitatively account for the salt concentration and $p$H dependence of the second virial coefficient of protein solutions, we now use DLVO theory to determine the effective hard-core diameter in a way that is independent of our one-parameter fitting approach (Sec.~\ref{sec:3}). 
If the solution conditions are known (Tab.~\ref{tab:1}), the temperature dependent effective hard-core diameter within the DLVO model $\sigma_\text{eff}^\text{DLVO}(T)$ can be computed based on Eq.~(\ref{eq:eff}) and the repulsive part of the interaction potential $U_\text{HS}(r) + U_\text{SC}(r)$. 
We have calculated $\sigma_\text{eff}^\text{DLVO}(T)$ in the relevant temperature window $0.9 \leq T / T_\text{c} \leq 1.2$. 
The results are shown for various solution conditions in Figure~\ref{fgr:7}a (lines). 
For comparison, we also show the temperature-independent values of $\sigma_\text{eff}$, as determined by our scaling approach in Sec.~\ref{sec:3} (symbols, at an arbitrary temperature). 
For the temperatures studied, the values of $\sigma_\text{eff}^\text{DLVO}$ lie in a very narrow range ($1.04 \leq \sigma_\text{eff}^\text{DLVO} / \sigma \lesssim 1.09$) and show only a very weak temperature dependence, whereas the values of $\sigma_\text{eff}$ are larger than those of $\sigma_\text{eff}^\text{DLVO}$ and lie in a broader range ($1.16 \leq \sigma_\text{eff} / \sigma \leq 1.34$). 
Nevertheless, the weak temperature dependence of $\sigma_\text{eff}^\text{DLVO}$ suggests that the assumed $T$-independence of $\sigma_\text{eff}$ represents a reasonable first approximation.

Having calculated the effective hard-core diameter within the DLVO model, $\sigma_\text{eff}^\text{DLVO}$, we have performed a similar analysis with the experimental binodals and $b_2(T)$ values as in Sec.~\ref{sec:4} and \ref{sec:5}, but based on $\sigma_\text{eff}^\text{DLVO}(T)$ instead of $\sigma_\text{eff}$. 
As the values of $\sigma_\text{eff}^\text{DLVO}$ are much lower than those of $\sigma_\text{eff}$, the protein binodals, when scaled to the critical temperature and the effective volume fraction $\phi_\text{eff}^\text{DLVO}$, do not collapse with the binodals of short-range SW fluids, but, for a given temperature, lie at smaller $\phi_\text{eff}$ (not shown). 
However, the $b_2^{\star\text{DLVO}}(T)$ data again lie near a master curve, which, following Eq.~(\ref{eq:4}), is shifted to more negative values (not shown). 
Based on these calculations, we present the experimental binodals in the $b_2^{\star\text{DLVO}} - \phi_\text{eff}^\text{DLVO}$ plane together with the grey shaded Noro-Frenkel ELCS area and simulation results of AHS and SW fluids (Fig.~\ref{fgr:7}b). 
As expected, the binodals in the $b_2^{\star\text{DLVO}} - \phi_\text{eff}^\text{DLVO}$ plane are, in general, lower and toward smaller $\phi_\text{eff}^\text{DLVO}$ than those obtained based on $\sigma_\text{eff}$ (Fig.~\ref{fgr:5}). 
They therefore tend to lie at or below the lower limit of the grey-shaded area of the ELCS mapping predictions. 

Thus, DLVO theory seems to consider only parts of the repulsive contributions to protein interactions (with respect to the ELCS mapping). 
A similar argument with respect to the repulsive interactions as quantified by DLVO theory has been proposed by Broide et al.\cite{Broide1996} and others.\cite{Israelachvili1996,Vekilov2007} 
They argued that DLVO based calculations are unable to explain the stability of protein solutions against aggregation.\cite{Broide1996} 
Furthermore, they suspect that repulsive protein hydration forces significantly contribute to protein--protein interactions. 
Whether a similar argument applies here needs to be examined in further experimental and theoretical studies. 

\section{Estimating second virial coefficients based on cloud-point measurements}\label{sec:7}
Based on our analysis of gas--liquid binodals (Sec.~\ref{sec:3}) and the temperature dependence of second virial coefficients of protein solutions (Sec.~\ref{sec:4}), we propose a method that allows for a quantitative estimation of $b_2$ values at a temperature $T$ near the critical temperature $T_\text{c}$ based on cloud-point measurements. 
To this end, it is necessary to measure cloud point temperatures at various protein concentrations at the respective solution condition, so that the critical temperature $T_\text{c}$ can be inferred based on Eq.~(\ref{eq:2}) and the effective hard-core diameter $\sigma_\text{eff}$ can be determined by our scaling approach (Sec.~\ref{sec:3}). 
If $T$ is not much smaller or larger than $T_\text{c}$ ($0.95 \leq T/T_\text{c} \leq 1.05$), possible deviations from the universal fit curve to $b_2^\star (T/T_\text{c})$ are expected to be small. 
Therefore, $b_2 (T/T_\text{c}) = \left( \sigma_\text{eff} / \sigma \right)^3 b_2^\star (T/T_\text{c})$ can be calculated if $T_\text{c}$ and $\sigma_\text{eff}$ are known. 
Following this approach, we have determined $b_2$ values at room temperature $T = 293.15~\text{K}$ for the binodals shown in Figure~\ref{fgr:3}c, which only contain NaCl as an additive.\cite{Muschol1997,Gibaud2011,Goegelein2012}
The results are shown in Figure~\ref{fgr:6} (crosses) as a function of salt concentration and $p$H value. 
In the high-salt regime ($c_\text{s} > 0.3~\text{M}$), i.e. the short-range limit, the results of the model calculations are in quantitative agreement with experimental data (symbols) as well as with the independent DLVO-based calculation of $b_2$ values (lines). 
We have tested our approach also for the other solution conditions of the experimental $b_2$ values discussed in this work (not shown);\cite{Goegelein2012,Gibaud2011} 
for the $T$ window given above, deviations between model calculation and experimental values are on the order of the experimental error, while for $T \gg T_\text{c}$ and for large values of $\sigma_\text{eff} / \sigma$, deviations are significant.

Our approach to calculate $b_2$ values, based on CPT measurements and a comparison with our SW fluid data, should also apply to other solution conditions, in particular, in the presence of additives and in the high-salt limit, where the experimental determination of $b_2$ values can be difficult.

\section{Concluding remarks}\label{sec:8}
In this work, we presented a direct and quantitative comparison of the experimentally observed binodals of protein (lysozyme) solutions and short-range square-well (SW) fluids. 
The temperature axis of the experimental binodals was scaled to the second virial coefficient and the concentration axis with the effective hard-core diameter which takes into account the repulsive contributions to the interactions.
For simplicity, we assumed that the effective hard-core diameter is temperature independent. 
It was determined by a comparison of experimental binodals of the protein solutions to our simulation binodals of SW fluids. 
Following this procedure, the experimental data sets quantitatively agreed with the binodals of short-range SW fluids and of the AHS fluid, corroborating the extended law of corresponding states as proposed by Noro and Frenkel.\cite{Noro2000} 
Thus, our results demonstrate the predictive power of the ELCS also for systems as complex as protein solutions.

If the effective hard-core diameter and hence the repulsive parts of the interactions are taken into account, the second virial coefficients of the protein solutions follow a master curve for temperatures close to the criticial temperature.
Based on this finding, we have proposed an approach to estimate second virial coefficients based on cloud-point measurements only. 
The obtained $b_2$ values for various salt concentrations and $p$H values are in quantitative agreement with results from scattering experiments and DLVO-based calculations. 

Our findings call for systematic experiments to study both the binodals of protein solutions and the temperature dependence of $b_2$ for further solution conditions, in particular different additives.

\section*{Appendix}
In Table~\ref{tab:2}, we list our new simulation data for SW fluids with $\lambda = 1.05$ and $1.1$.

\begin{table}
\caption{\label{tab:2} Monte Carlo simulation results for the gas--liquid binodal of SW fluids with $\lambda = 1.05$ (top) and $\lambda = 1.1$ (bottom) for various scaled temperatures $T^\star$ and densities of the gas and liquid phase, $\rho_\text{G}$ and $\rho_\text{L}$, respectively.}
\begin{ruledtabular}
\begin{tabular}{ccc}
$T^\star$ & $\rho_\text{G}$ &  $\rho_\text{L}$  \\ \hline
0.3575   & 0.13393 & 0.82045 \\
0.36   & 0.16147 & 0.78539 \\
0.3625   & 0.2047 & 0.7439 \\
0.365   & 0.27549 & 0.67303 \\
0.36679  & 0.4748 & 0.4748  \\ \hline
0.46   & 0.109 & 0.826 \\
0.465  & 0.1438 & 0.81346 \\
0.4675  & 0.15931 & 0.79708 \\
0.47   & 0.17854 & 0.76189 \\
0.4725  & 0.21401 & 0.73456 \\
0.475  & 0.248 & 0.687 \\
0.47272  & 0.47961 & 0.47961 \\
\end{tabular}
\end{ruledtabular}
\end{table}

\section*{Acknowledgement}
We thank D.~Wagner and H.~L\"owen (University of D\"{u}sseldorf, Germany) for very helpful discussions. 
We gratefully acknowledge support by the Deutsche Forschungsgemeinschaft (DFG). 
R.C.-P. acknowledges financial support provided by the Marcos Moshinsky fellowship 2013-2014.

\bibliography{guanidine_bib}
 
\end{document}